\documentclass[twocolumn]{aastex631}
\usepackage{amsmath}
\usepackage{upgreek}
\usepackage{CJK}

\newcommand{\uatnum}[1]{\href{http://vocabs.ands.org.au/repository/api/lda/aas/the-unified-astronomy-thesaurus/current/resource.html?uri=http://astrothesaurus.org/uat/#1}{#1}}
\newcommand{\caltech}{Department of Astronomy, California Institute of Technology, Pasadena, CA 91125, USA}
\newcommand{\gps}{Division of Geological \& Planetary Sciences, California Institute of Technology, Pasadena, CA 91125, USA}
\newcommand{\ucsc}{Department of Astronomy \& Astrophysics, University of California, Santa Cruz, CA95064, USA}
\newcommand{\keck}{W. M. Keck Observatory, 65-1120 Mamalahoa Hwy, Kamuela, HI, USA}
\newcommand{\ucla}{Department of Physics \& Astronomy, 430 Portola Plaza, University of California, Los Angeles, CA 90095, USA}
\newcommand{\jpl}{Jet Propulsion Laboratory, California Institute of Technology, 4800 Oak Grove Dr.,Pasadena, CA 91109, USA}
\newcommand{\ucsd}{Center for Astrophysics and Space Sciences, University of California, San Diego, La Jolla, CA 92093}
\newcommand{\ifahonolulu}{Institute for Astronomy, University of Hawai`i, 2680 Woodlawn Drive, Honolulu, HI 96822, USA}
\newcommand{\berkeley}{Department of Astronomy, University of California at Berkeley, CA 94720, USA}
\newcommand{\ames}{NASA Ames Research Centre, MS 245-3, Moffett Field, CA 94035, USA}
\newcommand{\water}{H$_2$O}
\newcommand{\methane}{CH$_4$}

\submitjournal{AAS Journals}

\shorttitle{HR 8799 Planets with KPIC}
\shortauthors{Wang et al.}

\begin{document}
\begin{CJK*}{UTF8}{gbsn}

\title{Detection and Bulk Properties of the HR 8799 Planets with High Resolution Spectroscopy}
\correspondingauthor{Jason J. Wang}
\email{jwang4@caltech.edu}

\author[0000-0003-0774-6502]{Jason J. Wang (王劲飞)}
\altaffiliation{51 Pegasi b Fellow}
\affiliation{\caltech}

\author[0000-0003-2233-4821]{Jean-Baptiste Ruffio}
\affiliation{\caltech}

\author{Evan Morris}
\affiliation{\ucsc}

\author[0000-0001-8953-1008]{Jacques-Robert Delorme}
\affiliation{\keck}
\affiliation{\caltech}

\author[0000-0001-5213-6207]{Nemanja Jovanovic}
\affiliation{\caltech}

\author{Jacklyn Pezzato}
\affiliation{\caltech}

\author{Daniel Echeverri}
\affiliation{\caltech}

\author[0000-0002-1392-0768]{Luke Finnerty}
\affiliation{\ucla}

\author{Callie Hood}
\affiliation{\ucsc}

\author{J. J. Zanazzi}
\affiliation{Canadian Institute for Theoretical Astrophysics, University of Toronto, 60 St. George Street, Toronto, ON M5S 3H8, Canada}

\author{Marta L. Bryan}
\altaffiliation{51 Pegasi b Fellow}
\affiliation{\berkeley}

\author{Charlotte Z. Bond}
\affiliation{\keck}

\author{Sylvain Cetre}
\affiliation{\keck}

\author[0000-0002-0618-5128]{Emily C. Martin}
\affiliation{\ucsc}

\author{Dimitri Mawet}
\affiliation{\caltech}
\affiliation{\jpl}

\author{Andy Skemer}
\affiliation{\ucsc}

\author{Ashley Baker}
\affiliation{\caltech}

\author{Jerry W. Xuan}
\affiliation{\caltech}

\author{J. Kent Wallace}
\affiliation{\jpl}

\author[0000-0002-4361-8885]{Ji Wang (王吉)}
\affiliation{Department of Astronomy, The Ohio State University, 100 W 18th Ave, Columbus, OH 43210 USA}


\author{Randall Bartos}
\affiliation{\jpl}

\author{Geoffrey A. Blake}
\affiliation{\gps}

\author{Andy Boden}
\affiliation{\caltech}

\author{Cam Buzard}
\affiliation{Division of Chemistry and Chemical Engineering, California Institute of Technology, Pasadena, CA 91125, USA}

\author{Benjamin Calvin}
\affiliation{\caltech}
\affiliation{\ucla}

\author{Mark Chun}
\affiliation{Institute for Astronomy, University of Hawai`i at M\={a}noa, 640 North A`ohoku Place, Hilo, HI 96720-2700, USA}

\author{Greg Doppmann}
\affiliation{\keck}

\author[0000-0001-9823-1445]{Trent J.~Dupuy}
\affiliation{Institute for Astronomy, University of Edinburgh, Royal Observatory, Blackford Hill, Edinburgh, EH9 3HJ, UK}

\author[0000-0002-5092-6464]{Gaspard Duch\^ene}
\affiliation{\berkeley}
\affiliation{Universit\'e Grenoble-Alpes, CNRS Institut de Plan\'etologie et d'Astrophysique (IPAG), F-38000 Grenoble, France}

\author{Y. Katherina Feng}
\affiliation{\ucsc}

\author[0000-0002-0176-8973]{Michael P. Fitzgerald}
\affiliation{\ucla}

\author{Jonathan Fortney}
\affiliation{\ucsc}

\author{Richard S. Freedman}
\affiliation{SETI Institute, Mountain View, CA 94043, USA}
\affiliation{\ames}

\author{Heather Knutson}
\affiliation{\gps}

\author{Quinn Konopacky}
\affiliation{\ucsd}

\author{Scott Lilley}
\affiliation{\keck}

\author{Michael C. Liu}
\affiliation{\ifahonolulu}

\author{Ronald Lopez}
\affiliation{\ucla}

\author{Roxana Lupu}
\affiliation{BAER Institute, NASA Research Park, Moffett Field, CA 94035, USA
}

\author{Mark S. Marley}
\affiliation{Department of Planetary Sciences and Lunar and Planetary Laboratory, University of Arizona, Tucson, AZ 85721, USA}

\author{Tiffany Meshkat}
\affiliation{IPAC, California Institute of Technology, M/C 100-22, 1200 East California Boulevard, Pasadena, CA 91125, USA}

\author{Brittany Miles}
\affiliation{\ucsc}

\author{Maxwell Millar-Blanchaer}
\affiliation{Department of Physics, University of California, Santa Barbara, Santa Barbara, California, USA}

\author{Sam Ragland}
\affiliation{\keck}

\author{Arpita Roy}
\affiliation{Space Telescope Science Institute, Baltimore, MD 21218, USA}

\author[0000-0003-4769-1665]{Garreth Ruane}
\affiliation{\caltech}
\affiliation{\jpl}

\author{Ben Sappey}
\affiliation{\ucsd}

\author{Tobias Schofield}
\affiliation{\caltech}

\author{Lauren Weiss}
\affiliation{\ifahonolulu}

\author{Edward Wetherell}
\affiliation{\keck}

\author{Peter Wizinowich}
\affiliation{\keck}

\author{Marie Ygouf}
\affiliation{NASA Exoplanet Science Institute, IPAC, Pasadena, CA 91125, USA}

\begin{abstract}
Using the Keck Planet Imager and Characterizer (KPIC), we obtained high-resolution (R$\sim$35,000) $K$-band spectra of the four planets orbiting HR 8799. We clearly detected \water{} and CO in the atmospheres of HR 8799 c, d, and e, and tentatively detected a combination of CO and \water{} in b. These are the most challenging directly imaged exoplanets that have been observed at high spectral resolution to date when considering both their angular separations and flux ratios. We developed a forward modeling framework that allows us to jointly fit the spectra of the planets and the diffracted starlight simultaneously in a likelihood-based approach and obtained posterior probabilities on their effective temperatures, surface gravities, radial velocities, and spins. We measured $v\sin(i)$ values of $10.1^{+2.8}_{-2.7}$~km/s for HR 8799 d and $15.0^{+2.3}_{-2.6}$~km/s for HR 8799 e, and placed an upper limit of $< 14$~km/s of HR 8799 c. Under two different assumptions of their obliquities, we found tentative evidence that rotation velocity is anti-correlated with companion mass, which could indicate that magnetic braking with a circumplanetary disk at early times is less efficient at spinning down lower mass planets. 
\end{abstract}


\section{Introduction}
In the past two decades, direct imaging has discovered several dozen substellar companions with masses from 1-70~$M_\textrm{Jup}$ with orbital separations from $\sim$3 au out to $\sim$1000~au. \citep[for a review, see][]{Bowler2016}.
The occurrence rates of giant planets and brown dwarfs beyond $\gtrsim 10$~au have begun to show that multiple formation channels are responsible for the current population of imaged substellar companions \citep{Nielsen2019, Vigan2020shine}. Between 10-100 au, \citet{Nielsen2019} showed that giant planets between 5-13~$M_\textrm{Jup}$ have a higher occurrence rate compared to their brown dwarf counterparts (13-80~$M_\textrm{Jup}$) and preferentially occur around higher mass stars, indicating the known exoplanet companions likely formed as the high-mass tail of planet formation through core accretion \citep{Pollack1996}, whereas brown dwarf companions formed like binary stars through gravitational instability \citep{Boss1998, Boss2001}. \citet{Vigan2020shine} found a similar dichotomy looking at the mass ratios between the companions and their host stars. They inferred that companions around lower mass stars with mass ratios closer to unity formed like binary stars whereas more extreme mass ratio companions around more massive stars formed like planets. 

Since the directly imaged companions are amenable to spectroscopy and astrometric monitoring, we can use population-level characteristics beyond detection to study this population and understand how they formed. \citet{Bowler2020} reinforced the finding of multiple formation channels by showing evidence that giant planets at wide separations (5-100~au) had an eccentricity distribution similar to that of close-in ($< 1$~au) giant planets, whereas the brown dwarf eccentricity distribution resembled the stellar-binary population. Measurements of the spin of planetary mass companions have pointed to magnetic braking quickly slowing down the spin rate of planets after formation \citep{Bryan2020}. While there has not yet been a population-level study of atmospheric compositions, compositional studies of individual objects are able to contribute evidence to discerning their formation channels \citep{Konopacky2013, Barman2015, GRAVITY2020, Molliere2020, Wilcomb2020}. 

High-resolution spectroscopy of directly imaged companions allows us to characterize their orbits, spin, and compositions. The Doppler shift of molecular absorption lines in the planetary atmosphere allows us to measure the radial velocity of the planet, which is useful to break degeneracies between orbital inclination and eccentricity for companions with limited orbital coverage \citep{Snellen2014, Schwarz2016, Ruffio2019}. The rotational broadening of these absorption lines allows for a direct measurement of planetary spin \citep{Snellen2014, Schwarz2016, Bryan2018}. The detection of molecular signatures through cross-correlation methods takes advantage of the fact the planet and star have different spectral features, enables the detection of trace molecular species, and allows for the inference of planetary composition \citep{Konopacky2013, Barman2015, Brogi2019, Wilcomb2020}.

However, up until now, slit spectrographs assisted with adaptive-optics (AO) have been observationally limited to bright companions that are at relatively large angular separations from their host stars. For planetary-mass companions within one arcsecond, only $\beta$ Pic b and HR 8799 c have been detected at a spectral resolution $R > 10,000$ \citep{Snellen2014, WangJi2018}. The main difficulty is that the bright glare of the host star overwhelms the signal of the planet. The glare of the star can be mitigated by combining high-contrast imaging techniques with high-resolution spectroscopy \citep{Snellen2015}. The combinations of these two techniques, which we term high dispersion coronagraph (HDC), drives the design of the Keck Planet Imager and Characterizer \citep[KPIC;][Delorme et al. submitted]{Mawet2017,Jovanovic2019}. KPIC combines starlight suppression using the Keck AO system and single mode fibers with the NIRSPEC high-resolution spectrograph to enable high-resolution spectroscopy of fainter and closer-in directly imaged planets. Similar instrument designs are also being pursued by Subaru/REACH \citep{jovanovic2017_AO4ELT,Kotani2020} and VLT/HiRISE \citep{Vigan2018,Otten2021}.

In this paper, we present the first science demonstration of KPIC with observations of the four planets orbiting HR 8799. HR 8799 is a notable planetary system, as it is the only known system with four directly imaged exoplanets \citep{Marois2008,Marois2010}. The four planets are either near or in mean-motion resonance, and dynamical modeling of their orbits have constrained their masses to be $7.2 \pm 0.7$~$M_\textrm{Jup}$ for the inner three planets and $5.8 \pm 0.5 $~$M_\textrm{Jup}$ for planet b \citep[][also see \citealt{Gozdziewski2020}]{Wang2018}. Since their discovery, the planets have been characterized extensively in the 1-5~$\upmu$m range using broadband photometry and low- and medium-resolution spectroscopy \citep{Bowler2010, Barman2011, Konopacky2013, Currie2014, Ingraham2014, Skemer2014, Zurlo2016, bonnefoy2016, Greenbaum2018, Molliere2020}. At medium-resolution, the individual molecular lines begin to be resolved spectrally, allowing for detection of molecular signatures as well as the radial velocity of the planet \citep{Barman2011, Konopacky2013, Ruffio2019}. Measurements of elemental abundances in these planetary atmospheres have shown deviations from the stellar abundances, which have been interpreted to mean that these planets formed via core accretion rather than gravitational instability \citep{Konopacky2013, Barman2015, Lavie2017, Molliere2020}. 
 
Section \ref{sec:obs} details the observations of the HR 8799 planets made with KPIC. Section \ref{sec:data-redu} describes the initial data reduction steps. The detection of molecular features through cross correlation as well as fitting atmospheric models of rotating planets directly to the data is discussed in Section \ref{sec:fits}. We obtained the first spin measurements for these exoplanets, and we put them, as well as our measured orbital velocities and bulk atmospheric properties, in context in Section \ref{sec:discussion}. We summarize our work and discuss future avenues both in obtaining better data and utilizing better models to study these planets in Section \ref{sec:conc}. 

\section{Observations}\label{sec:obs}
\subsection{Instrument Description}

\begin{figure*}
    \centering
    \includegraphics[width=0.8\textwidth]{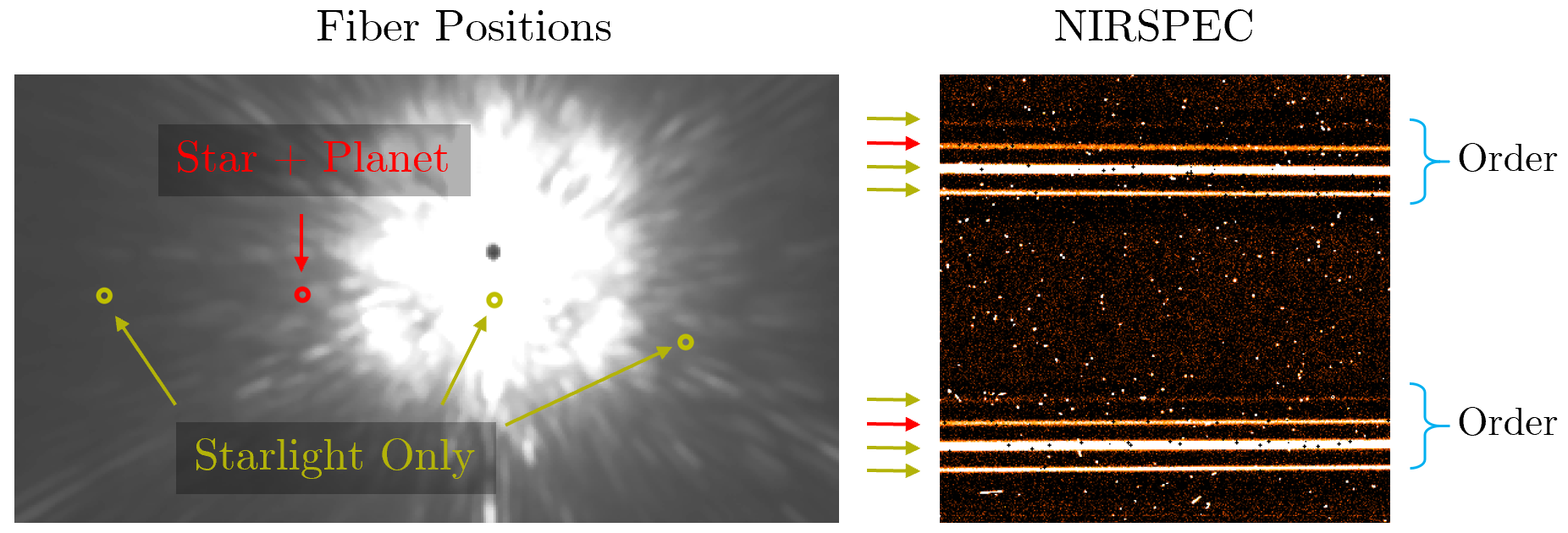}
    \caption{Schematic of KPIC observations of the HR 8799 planets. On the left are the locations of the fibers relative to a Keck/NIRC2 PSF of the star for an observing sequence on HR 8799 c. The red circle marks the location of the fiber centered on the planet, while the yellow circles mark the positions of the three other fibers that were in use. A fifth fiber is not shown in this image as it was blocked by the NIRSPEC slit and was not used. Note that the NIRC2 image was not taken during the observing sequence and is just to serve as an example to show the dynamic range of the stellar halo. On the right is a small portion ($\sim$360x370 pixels) of the NIRSPEC detector image in this configuration. A portion of two echelle orders are shown. In each order, four spectral traces are seen corresponding to the placement of the four fibers shown on the left image. }
    \label{fig:schematic}
\end{figure*}

The Keck Planet Imager and Characterizer (KPIC) consists of a series of upgrades for the Keck II AO system and its two facility instruments: the NIRC2 infrared imager and the upgraded NIRSPEC infrared high-resolution spectrograph \citep{McLean1998, Martin2018, Lopez2020}. As part of this project, instrument upgrades included a new vortex coronagraph for NIRC2 \citep{Vargas2016,Serabyn2017} and an infrared pyramid wavefront sensor operating in H band for the AO system \citep{Bond2020}. Particularly relevant for high resolution spectroscopy of directly imaged exoplanets, a fiber injection unit (FIU) following the concept presented in \citet{Mawet2017} was deployed in 2018. The FIU benefits from the NIRSPEC detector upgrade that allows KPIC to reach $R\sim35,000$ \citep{Martin2014, Martin2018, Lopez2020}. We point the reader to Delorme et al. (submitted) for detailed information about the instrumentation. 

Here, we provide a brief summary of the relevant instrumentation. The pyramid wavefront sensor drives the facility deformable mirrors in the AO system to compensate for atmospheric turbulence. In addition to imaging exoplanets using NIRC2, KPIC can also send the $K$-band light of the system to the FIU to spectroscopically characterize these planets. Located after the AO system, the FIU steers the light of a planet into one of five single mode fibers represented by circles on Figure~\ref{fig:schematic}. These fibers are connected to NIRSPEC to spectrally disperse the light injected into the fibers (see right panel of Figure~\ref{fig:schematic}). Since the star is bright, diffracted starlight (stellar ``speckles") leaks into all of the fibers, with the amount of starlight in each fiber depending on the angular distance between the fiber position on the sky and the star. In \citet{Mawet2017_ApJ}, we demonstrated that the use of single-mode fibers provide the following advantages: a well-defined Gaussian-like line spread function (LSF) which is independent in shape to input wavefront aberrations, and, on average, 3x suppression of the underlying stellar speckle flux at the location of the fiber. Note that the current KPIC FIU does not utilize a coronagraph to suppress diffracted starlight, so the utilization of adaptive optics with the single mode fibers is the main starlight suppression mechanism.

The fiber extraction unit (FEU) reimages the light from the single mode fibers onto the NIRSPEC spectrograph slit, where the light is then dispersed inside the spectrograph. The resulting NIRSPEC data has up to five fibers illuminated with light in each echelle order (only four fibers were imaged onto the detector for the HR 8799 observations). Each fiber roughly subtends an angular diameter of $\sim50$~mas. The projected separation between two consecutive fibers is $\sim$800 mas for the four fibers shown. Because the slit sees the FEU and not the sky, the slit background is dominated primarily by the thermal emission from the warm room-temperature optics of the FEU, with a lesser contribution from the thermal instrument background of NIRSPEC. The sky background only appears in the fibers, but is generally well below the thermal background of the FEU, with only weak OH line emission seen in long exposures.

\subsection{HR 8799 Observations}

\begin{deluxetable*}{c|c|c|c|c}
\tablecaption{HR 8799 Planet Observations \label{table:obs}}
\tablehead{ 
Date & Target & Integration Time (min) & Airmass & Throughput (\%)
}
\startdata 
2020-07-01 & HR 8799 c & 230 & 1.0 - 1.8  & 1.7 - 1.9\\
2020-07-02 & HR 8799 d & 230 & 1.0 - 1.7  & 1.9 - 2.4\\
2020-07-03 & HR 8799 e & 130 & 1.1 - 2.0  & 2.2 - 2.4\\
2020-09-29 & HR 8799 b & 160 & 1.0 - 1.1  & 1.8 - 2.4\\
\enddata
\end{deluxetable*}

We listed the epochs of observations for each planet in Table \ref{table:obs}. Each planet was observed with a similar observing sequence. NIRSPEC was set up to use the ``Thin" filter, a thin piece of clear PK50 glass which blocks light at wavelengths longer than $\sim2.5$ $\upmu$m, along with a custom pupil stop designed for the KPIC FEU and the NIRSPEC 0.0679"x1.13" slit to maximize the signal to noise of the light from each fiber relative to instrument background. The NIRSPEC echelle grating was set to $63.0^\circ$ and the cross disperser was set to $35.76^\circ$ to allow us to obtain nine spectral orders ranging from approximately 1.94 to 2.49 $\upmu$m. 

We started each observing sequence by placing the host star on each fiber and took at least one exposure (30-60~s) of the host star in all four fibers for data calibration purposes. We then designated one of the fibers as the primary science fiber, based on the fiber that had the best throughput in daytime testing. We used the tip/tilt mirror in the FIU to offset the star from the fiber bundle such that the planet of interest is placed on the primary science fiber. The science fiber received a combination of planet light plus diffracted residual starlight. The offset amplitude and direction was computed using the \texttt{whereistheplanet}\footnote{http://whereistheplanet.com/} orbit prediction tool \citep{whereistheplanet} based on the dynamically stable and coplanar orbit fit from \citet{Wang2018}. From preliminary instrument characterization efforts, we estimated that the offset accuracy is 10~mas, which corresponds to $< 10\%$ loss in throughput due to fiber misalignment. Although the fibers in the bundle were fixed in a linear arrangement relative to one another, the adaptive optics field rotator (K-mirror) was used to rotate the field-of-view such that the star was coupled into as many of the other fibers as possible so that we could obtain simultaneous stellar spectra. 

We then took 600~s exposures with NIRSPEC using the MCDS-16 detector readout mode in this configuration where one fiber had the light of the planet and at least two other fibers were transporting significant amounts of starlight.This integration time was chosen to be long enough so that read noise is negligible (thermal background noise $\sim$10x larger), but short enough to tolerate any bad frames due to technical issues such as the AO loops opening. Every hour or so, we placed the host star on each science fiber and took a short exposure (30-60~s) for calibration. The open-shutter time obtained on each planet is listed in Table~\ref{table:obs}.

Using the short exposures on the star, we calculated the end-to-end throughput from the top of the atmosphere to the detector. We reported the end-to-end throughput measured between 2.29-2.34~$\upmu$m in Table~\ref{table:obs} as a metric that combines both instrument performance and atmospheric conditions during the observations. We note that this wavelength range is not the best performing wavelength, as Delorme et al. (submitted) showed that a throughput of over 3\% has been achieved at shorter wavelengths. However, our data analysis in the following sections is focused around this wavelength range since it coincides with the CO bandhead in $K$-band, so this is the most relevant throughput metric. Overall, we see that the conditions between the four nights are pretty comparable, with 2020-07-01 having slightly poorer performance due to issues in fiber injection that were fixed after the night. 

\subsection{Calibration Data}
In addition to the spectra of the star we obtained during the observing sequence, we also observed an M giant (HIP 81497) for wavelength solution calibration and a telluric standard star for the wavelength calibrator (ups Her) each night. We took five 1.5~s exposures of HIP 81497 and three 30~s exposures of ups Her on-axis in each of the four fibers. After the night, we took thermal background frames looking at the FEU with no light injected into the fibers at each of the exposure times used to model the thermal background of the instrument.

\section{Data Reduction}\label{sec:data-redu}

\subsection{Raw Data Reduction}\label{sec:data-raw}
The process of going from the original detector images to extracted 1D spectra was the same for all data regardless of the object being observed. First, the images were background subtracted using the instrument thermal background frames taken when no light was being injected into the fibers during the daytime. The FEU typically was at a different temperature during the day so these thermal frames did not perfectly subtract the data and leave some residual background which we modeled during the extraction step. Thermal images taken during the observing sequence would have provided better background subtraction (i.e., nodding), but were not acquired as we had not developed an efficient way to nod the planet light on the detector.

For each night of observation, the trace of each of the four science fibers in each of the nine orders was determined by using the data on the telluric standard star ups Her. As the point spread function (PSF) of monochromatic light coming from a single mode fiber is nearly a 2D Gaussian, we just needed to measure the position and standard deviation of the PSF at each wavelength. To do that, we fit a 1D Gaussian to the cross section of the trace in each column of each order to determine the position and standard deviation of the Gaussian PSF at that column. For each of the nine orders, we recorded the position and standard deviation of the Gaussian PSF for each of the four fibers in each column of the detector (2048 in total). To mitigate measurement noise and biases from telluric lines, we smoothed the measurements by fitting a cubic spline to the measured positions and standard deviations. These PSF standard deviations will also be used for estimating the line spread function (LSF) width in sections \ref{sec:xcorr} and \ref{sec:fm_fits}.

For each exposure, we then extracted the 1D spectra of each of the four fibers in each column of each order. Due to the imperfect background subtraction, we used pixels that are at least 5 pixels away from the center of any fiber to estimate the residual background level in each column and subtracted the median of these pixels from every pixel in the column. Then, for each fiber, we used optimal extraction to measure the flux using a 1D Gaussian profile as the PSF with the positions and standard deviations fixed to the values measured on the telluric standard star \citep{Horne1986}. The total integrated flux of the 1D Gaussian is then the flux we extracted for that fiber in that column. The uncertainty in the optimal extraction was used as the uncertainty in the flux measurement \citep{Horne1986}. 

\subsection{Wavelength Calibration}
We observed the M-giant HIP 81497 in each fiber each night to determine the wavelength solution from the stellar and telluric spectral lines. The wavelength solution was modeled as a spline using 6 nodes per order (\textit{i.e.}, piecewise-3\textsuperscript{rd} order polynomials). We modeled the data as the multiplication of a stellar spectrum, a telluric transmission term, and the transmission of the telescope and the instrument, complemented with an additive linear background term.
The star HIP 81497, with a M2.5III spectral type, was chosen from the Gaia RV standard catalog \citep{Soubiran2018}. It was modeled with a PHOENIX stellar spectrum \citep{Husser2013} assuming a temperature of $3600\,\textrm{K}$, surface gravity of $\textrm{log}(g)=1$, solar metallicity, and a fixed known RV of $-55.567\pm0.0011\,\mathrm{km/s}$. 
The telluric transmission of the Earth's atmosphere was modeled from a linearly interpolated grid of 25 ATRAN\footnote{\url{https://atran.arc.nasa.gov/cgi-bin/atran/atran.cgi}} models \citep{Lord1992} over water vapor overburden (500, 1000, 5000, 10000, and 20000 $\upmu$m) and zenith angle (0, 25, 45, 68, and 89 degrees). The water vapor overburden and the zenith angle were fit as nuisance parameters.
The transmission of the telescope and instrument varies with wavelength primarily due to the efficiency of the spectrograph to disperse light at each wavelength (i.e., blaze function). To model the spectrally dependent transmission, we used a piecewise-linear function that divided each order in 5 pieces.
The best fit wavelength solution was found using the Nelder-Mead optimization implemented in the \texttt{scipy.optimize.minimize} routine that jointly fit for the wavelength solution, telluric parameters, and the instrument and telescope transmission \citep{2020SciPy-NMeth}. The search was initialized around the minimum of a coarse grid search to avoid local minima.
This technique was found to be precise to the 0.1~km/s level~\citep{Morris2020}, which will be sufficient for the following data reduction. \citet{Morris2020} also found that the wavelength solution was stable within a night to the same level of precision as long as optics inside the spectrograph were not moved.

\section{Fitting High-Resolution Spectra}\label{sec:fits}
\subsection{Forward Modeling the Planet Spectra}
In this work, we build up a full forward model of the spectrum we recorded, which consists of both planet plus star light and is similar to the approach from \citet{Ruffio2019} for fitting medium-resolution integral field spectroscopy data of imaged exoplanets. The signal obtained from the fiber placed on one of the planets can be deconstructed into the following components:

\begin{equation}
    D_p(\lambda) = \alpha_p(\lambda) T(\lambda)P_{LSF}(\lambda) + \alpha_s(\lambda) T(\lambda)S_{LSF}(\lambda) + n(\lambda).
\end{equation}
Here, $D_p$ is the signal from the planet fiber, $T$ is the transmission of the optical system (atmosphere, telescope, and instrument) excluding fiber coupling efficiency, $P_{LSF}$ is the spectrum from the planet after it has been convolved by the instrumental LSF, $\alpha_p$ is a scaling term for the planet brightness accounting for fiber coupling efficiency, $S_{LSF}$ is the spectrum from the star after it is convolved by the instrumental LSF (this will be modeled by the empirical spectra of the star), $\alpha_s$ is a scaling term for the brightness of stellar speckles accounting for its fiber coupling efficiency, and $n$ is the noise (instrumental thermal background noise generally dominates). For simplicity, we will drop the $(\lambda)$ notation in the rest of the paper, but we note that all of these parameters will remain wavelength-dependent implicitly. 

The transmission of the optical system was calculated using on-axis observations of the star HR 8799. In these observations, the signal $D_s$ can simply be written as:
\begin{equation}
    D_s = T S_{LSF} + n
\end{equation}
We approximated $S_{LSF}$ using a model PHOENIX spectrum with an effective temperature of 7200~K \citep{Husser2013}. Given that HR 8799 is a F0 star \citep{Gray2003}, it has nearly no spectral lines in $K$-band, which mitigates any errors due to an imperfect stellar spectrum. Additionally, we did not fit the data near the Br$\gamma$ line in our analysis. The signal-to-noise ratio (SNR) per spectral channel of the stellar spectra is $\sim$400, so $n$ was assumed to be negligible. The transmission is then simply obtained by dividing the data by the model of the star: $T = D_s/S_{LSF}$. We used $T$ solely to model the transmission of the planet light from the top of the atmospheres to the detector. To model the stellar light that leaked into the planet fiber, we simply used the on-axis observations of the star, $D_s$, to account for $T(\lambda) \times S(\lambda)$ simultaneously. This has the added benefit that the exact shape of the LSF only matters for the planet signal, and is not used in fitting the stellar spectrum. 

Thus, we constructed a model for the data obtained on the planet fiber ($M_p$) as:
\begin{equation}\label{eq:model}
    M_p = \alpha_p T P_{LSF} + \alpha_s D_s
\end{equation}
where we can use planetary atmosphere models for $P_{LSF}$ to find the model atmosphere parameters that best fit the data. 

The terms $\alpha_p$ and $\alpha_s$ vary slowly as a function of wavelength. The wavelength dependence of $\alpha_p$ is dominated by differential atmospheric refraction changing the apparent sky position of the planet, and this effect changes slowly as a function of wavelength. The use of an atmospheric dispersion corrector in KPIC Phase II will mitigate this effect on $\alpha_p$ \citep{Wang2020SPIE,Jovanovic2020}. We did not find chromatic optical aberrations in the system to be measurable for inclusion in $\alpha_p$. 
On similar observations of a speckle field with a single mode fiber, \citet{GRAVITY2020} showed that $\alpha_s$ can be approximated across the entire $K$-band as a low-order polynomial. In some preliminary analysis to characterize $\alpha_p$ and $\alpha_s$ from observations of standard stars, we confirmed this and found that they change on $\sim$0.01~$\mu$m scales ($\sim$400 spectral channels).
This means that high-pass filtering the data would be adequate for continuum subtraction and will remove chromatic modulations in the continuum due to $\alpha_p$ and $\alpha_s$. 

To do this high-pass filter, we median filtered the spectrum with a 200-pixel ($\sim$0.004~$\upmu$m) box for computing the moving median and subtracted the median-filtered spectrum off from the original spectrum.
We note that this continuum subtraction does not account for the fact that $\alpha_p$ and $\alpha_s$ also induce chromatic modulations in the line depths, which we will ignore in this work as it is a smaller effect. In principle, this effect can be modeled in our KPIC data using a low-pass filter approach discussed in \citet{Ruffio2019}. 

We applied the high-pass filtering to both the data and model spectra. The model spectrum after high pass filtering can be written as:
\begin{equation}\label{eq:hpmodel}
    \mathcal{H}[M_p] \approx \mathcal{H}[\bar{\alpha}_p T P_{LSF} + \bar{\alpha}_s D_s],
\end{equation}
where $\mathcal{H}$ denotes high pass filtering, and $\bar{\alpha}_p$ and $\bar{\alpha}_s$ are wavelength independent versions of $\alpha_p$ and $\alpha_s$. They represent achromatic scaling terms for the mean planetary and stellar speckle flux levels in the data. The high-pass filter of the data would be written in the same way. 

In Section \ref{sec:xcorr} (but not in Section \ref{sec:fm_fits}), we made one further approximation assuming that the cross-talk between the various components is negligible when high-pass filtering using a median-filter implementation, allowing us to break down the model to filter the individual components:
\begin{equation}\label{eq:xcorr-data}
    \mathcal{H}[M_p] \approx \bar{\alpha}_p \mathcal{H}[ T P_{LSF}] + \bar{\alpha}_s \mathcal{H}[ D_s ]
\end{equation}
We pulled $\bar{\alpha}_p$ and $\bar{\alpha}_s$ outside of $\mathcal{H}$ since we have assumed they are constant values after we removed the slowly varying chromatic continuum with the high-pass filter.  We note that a linear implementation of the high-pass filter such as the Fourier-based approach from \citet{Ruffio2019} would make Equation \ref{eq:xcorr-data} exact. However, we found that a median-filter implementation modeled the continuum better than the Fourier-based ones we tried (simple cutoff in frequency or a Gaussian filter in frequency space). Removing the chromatic continuum of $\alpha_p$ and $\alpha_s$ was more important than ensuring perfect linearity in the high-pass filter. 

\subsection{Detection of Molecules}\label{sec:xcorr}

\begin{figure*}
    \centering
    \includegraphics[width=0.9\textwidth]{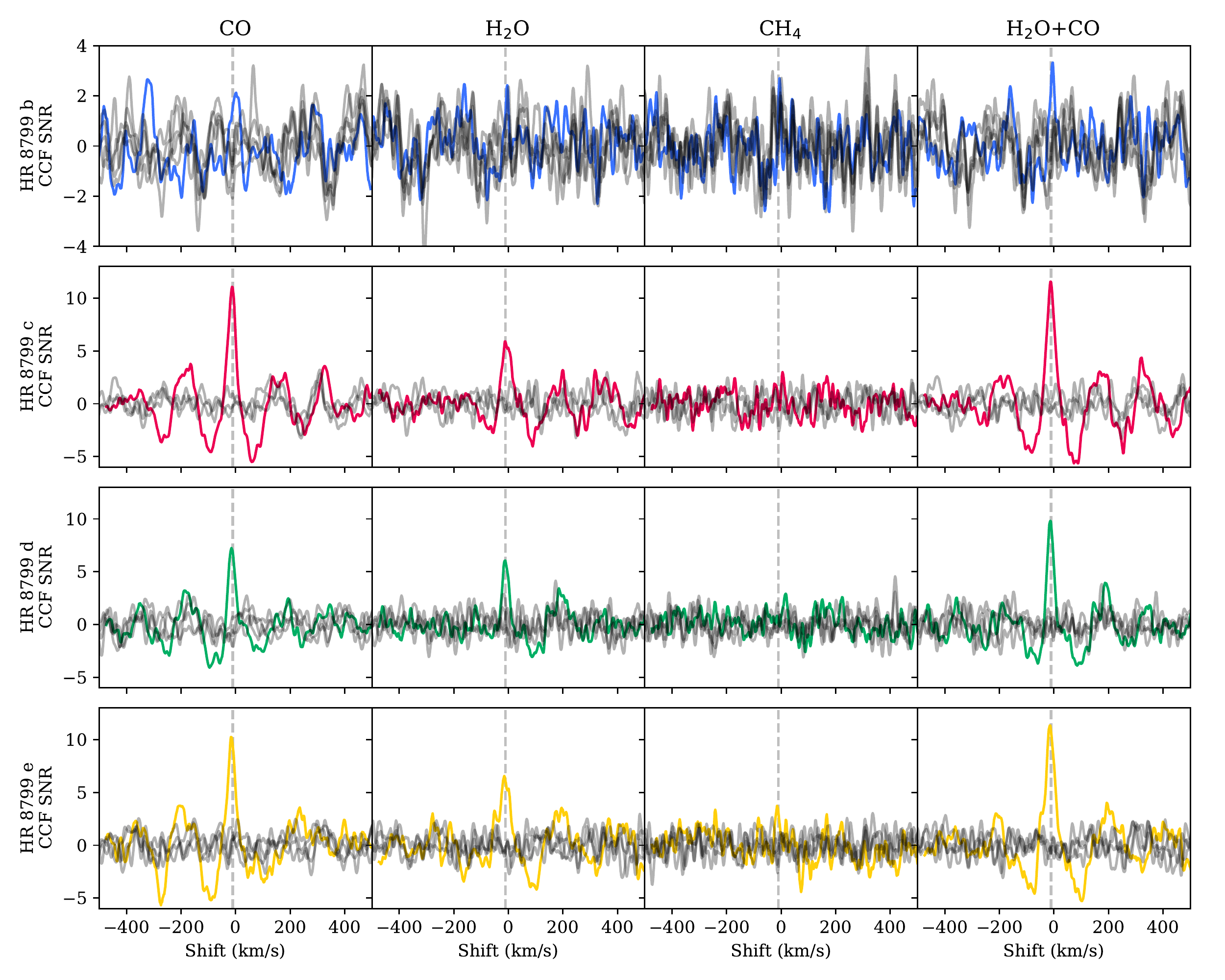}
    \caption{Cross correlation functions with individual molecular species for all four HR 8799 planets using the technique described in Section \ref{sec:xcorr}. For each planet's spectrum (one planet per row), the SNR of the CCF are plotted for molecules templates using CO, \water{}, \methane{}, \water{}+CO (one per column). The solid color line is the cross correlation function of each molecular template with the data on the planet. The solid gray lines are the cross correlation function of each template with extracted spectra that contained no planet signal. The vertical dashed line marks the approximate systemic radial velocity as calculated by \citet{Ruffio2019}. HR 8799 c, d, e have strong detections of CO and \water{} individually, whereas HR 8799 b is only weakly detected when combining \water{} and CO molecular templates.}
    \label{fig:xcorr}
\end{figure*}

First, we assessed the detection of all four planets by looking for the signature of molecular absorption lines from their planetary atmospheres as is common to do in the literature \citep[e.g.,][]{Konopacky2013, Ruffio2019, Xuan2020}. As shown in \citet{Ruffio2019thesis}, cross correlation techniques estimate the maximum likelihood value for the planet flux as a function of radial velocity (RV) shift for a given planet template $P_{LSF}$. As we have spatially resolved data that allows us to construct the planet spectrum and stellar spectrum separately, we performed a modified cross-correlation like in \citet{Ruffio2019thesis} where we estimated the maximum likelihood value for both the planet and star flux as a function of RV shift for a given planet template. For a given RV shift, we can rewrite Equation \ref{eq:xcorr-data} in matrix form to solve for the planet and star flux to best match the high-pass-filtered data $\mathcal{H}[D_p]$:
\begin{equation}
\begin{pmatrix}
\vdots  \\
\mathcal{H}[D_p] \\
\vdots 
\end{pmatrix}
= \begin{pmatrix}
\vdots  & \vdots  \\
\mathcal{H}[T P_{LSF}(RV)] & \mathcal{H}[D_s] \\
\vdots  & \vdots
\end{pmatrix}
\begin{pmatrix}
\bar{\alpha}_p \\
\bar{\alpha}_s 
\end{pmatrix}.
\end{equation}
Here, the data on the left hand side is a column vector with a length equal to the number of spectral channels, $N_\lambda$. The model on right hand side consists of a $N_\lambda \times 2$ matrix where the first column is the model flux of the planet and the second column is the model flux of the star (which we modeled empirically with the on-axis stellar spectrum), and a column vector of length 2 corresponding to the flux of the planet and star. The two unknowns are $\bar{\alpha}_p$ and $\bar{\alpha}_s$, making it straightforward to calculate their maximum likelihood values through linear least-squares optimization techniques. The cross correlation function (CCF) for a given planet model $P_{LSF}$ is then the calculated values of $\bar{\alpha}_p$ at every requested RV. Note that we recomputed $P_{LSF}$ at every RV shift considered. When combining data from multiple echelle orders, the data and model columns can simply be extended so that $N_\lambda$ is the total number of spectral channels from all of the orders considered in the fit. While not included in this analysis, each data point can also be weighted by its respective error when estimating $\bar{\alpha}_p$ by using equation D6 of \citet{Ruffio2019} for a more optimal ``matched filtering" detection metric. In the following analysis in this section, such error weighting made negligible changes to the CCF ($< 10$\%), so we have presented values without using it for simplicity.

For $P_{LSF}$, we generated molecular templates from the cloudless Sonora-Bobcat model set \citep[][Marley et al. submitted]{Marley2018}. Although this model lacked clouds, we found that the CCF signal from the cloudless Sonora-Bobcat CO+\water{} template was within 10\% of the CCF signal from the cloudy BT-Settl models used in Section \ref{sec:fm_fits} that contained all molecular opacities, indicating that the strength of our molecular detections does not strongly depend on cloud assumptions. Note that molecular templates constructed from the BT-Settl models are not available. We used the same atmospheric model parameters for all four planets: an effective temperature of 1200~K and a surface gravity of $\log(g) = 4.0$ based on the latest values for HR 8799 e from \citet{Molliere2020}. The detections were relatively insensitive to the exact choice of effective temperature and surface gravity: using a 1400~K model, which is representative of the scatter in effective temperature between different models \citep{Greenbaum2018}, changed the CCF signal by 5\%\footnote{Note, however, that the change in the likelihood of a model also depends on the number of data points so a small change in CCF signal can significantly alter the likelihood of a model \citep{Brogi2019}.}. Using the temperature structure and molecular abundance profile from this Sonora-Bobcat model and following the procedure in \citet{Morley2015}, we post-processed the atmosphere model profile to compute emergent spectra at $R = 200,000$, repeating the process to only consider the opacity of the molecule or molecules of interest each time. The model opacities used were from \citet{Freedman2008,Freedman2014}, which utilized the \methane{} line lists from \citet{Yurchenko2014} and \water{} line lists from \citet{Barber2006}. The templates were then convolved to instrumental resolution assuming the spatial size of the fiber trace measured in Section \ref{sec:data-raw} is equivalent to the width of the LSF. 

We computed CCFs using CO, \water, \methane, and CO+\water{} molecular templates. Spectra of each planet from echelle orders 31 (2.44-2.49 $\upmu$m), 32 (2.36-2.41 $\upmu$m), and 33 (2.29-2.34 $\upmu$m) were used, as these three orders had the best calibration. Including additional orders of the data or additional opacity sources in the model templates did not significantly increase the CCF signal. For each planet, we also computed CCFs for ``noise" spectra that were taken contemporaneously. These noise spectra were either extracted from other fibers with similar amounts of stellar flux leaking into them as the science fiber but were not pointed at the planet, or from regions of the detector that imaged the slit without any fibers and thus were dominated by thermal background noise. The CCFs were computed with velocity offsets between -500 and 500 km/s from the Solar System barycenter. The CCFs for each planet were normalized by dividing by the standard deviation of the CCFs from all the noise spectra, resulting in CCF signal-to-noise (CCF SNR) functions. The CCF SNR functions for each planet and for each molecular species are plotted in Figure~\ref{fig:xcorr}. We note that the CCFs of the noise spectra may not be Gaussian-distributed, so there maybe biases when quantifying the false alarm probabilities of these molecular detections.

We found strong detections of CO for HR 8799 c, d, and e with SNR between 7-11. For these three planets, we also have a strong detection of \water{} with SNR $> 5$. For HR 8799 b, we have a weak detection with SNR $\approx 3$ when combining the CO and \water{} templates, but there is no significant detection of any individual molecule alone. This is likely due to the fact HR 8799 b is about twice as faint as the other planets and had one of the shortest exposure times. Longer exposure times are needed to obtain confident detections of molecular signatures in HR 8799 b. For all four planets we did not detect \methane{} at a significant level. The non-detection of \methane{} despite the strong CO and \water{} detections are consistent with molecular detections of these planets at medium resolution with OSIRIS \citep[][Ruffio et al. submitted]{Konopacky2013, Barman2015, petit2018}. Our KPIC detections are the first detections of HR 8799 b, d, and e at high spectral resolution ($R > 10,000$), where many molecular absorption features start becoming spectrally resolved. Our 6$\sigma$ detection of \water{} in HR 8799 c in $K$ band is better than the previous 4.6$\sigma$ $L$-band detection of \water{} by NIRSPAO that used 3.5$\times$ more integration time \citep{WangJi2018}. 

\subsection{Bayesian Inference of Planetary Parameters}\label{sec:fm_fits}
We put our forward modeling approach into a Bayesian inference framework to fit our extracted spectra directly, retrieve atmospheric parameters, and assess how complex of a model is required to fit the data. Our likelihood-based framework is similar to the one developed by \citet{Ruffio2019} for medium resolution spectroscopy of the HR 8799 planets, although we did not analytically marginalize over any of the parameters in our likelihood. Unlike \citet{Brogi2019}, we did not use the cross correlation function in the likelihood, as we did not assume that each spectral channel has the same noise level. Our method is similar to the method used by \citet{Gibson2020} to characterize the Fe absorption on an ultra-hot Jupiter, except we simultaneously fit the star and planet together, which minimizes over-fitting of the planet signal when subtracting off the star using techniques such as principal component analysis \citep{Pueyo2016, Ruffio2019}. We further note that the relative flux ratio between the planet and the stellar components in the data are much less extreme in our case, which likely makes it easier to fit both components to the data simultaneously. 

We were interested in constraining the atmospheric and bulk parameters of the planets. For the planet spectrum, we fit the BT-Settl-CIFIST model grid to the data, varying both effective temperature ($T_\textrm{eff}$) and surface gravity ($\log(g)$ in cgs units) \citep{Allard2012}. We chose the BT-Settl models as it was the only publicly available grid of models that are available at a spectral resolution $R > 35,000$ and includes clouds, which have been shown to be important shaping these planets' spectra \citep[e.g.,][]{Molliere2020}. In particular for our high resolution spectra, the depths of molecular absorption lines change when clouds are included \citep{Hood2020}. Rather than just stepping through RV shifts as in Section \ref{sec:xcorr}, we fit for the planetary radial velocities relative to the Solar System barycenter. Note that the stellar radial velocity is not well determined making relative RV measurements challenging \citep{Ruffio2019}. Similar to Section \ref{sec:xcorr}, we can fit for the planet fluxes, which controls the depth of the planetary atmosphere lines compared to the stellar and telluric line depths. This can also be directly translated to $K$-band flux ratios of the planets which we can check against literature photometric measurements. We also fit for the rotational broadening of the planetary spectra ($v\sin(i)$) by using the \texttt{fastRotBroad} function in \texttt{PyAstronomy} to broaden our planet atmosphere models \citep{pyastronomy}.

We also improved on the stellar model by considering multiple $D_s$ exposures. In Section \ref{sec:xcorr}, we averaged all of the on-axis images of the star in time. Here, we fit a set of linear coefficients $c_i$ to optimally weigh the individual on-axis exposures of the star, $D_{s,i}$, to create the master stellar spectrum that best fits the data:
\begin{equation}
    D_s = \sum_i c_i D_{s,i} / \sum_i c_i.
\end{equation}
We note that this is similar to the LOCI technique in high-contrast imaging \citep{Lafreniere2007}, except that we optimized the coefficients while simultaneously fitting the planet model to the data like what has been done in medium-resolution spectroscopy \citep{Ruffio2019}. In our fits, we assumed the $c_i$ coefficients to be unchanged across orders, but the overall spectrum $D_s$ can be scaled by a different flux value to account for the chromaticity of the stellar speckle flux. 

Accurate estimation of errors is required for any robust statistical analysis. Preliminary analysis of the data indicated that the residuals to the forward model fits are dominated by uncorrelated noise. However, we found that the amplitude of the uncorrelated noise was higher than what the formal extraction errors predicted. This may be due to an underestimation of extraction errors or due to unaccounted for noise terms by $\sim$30\%. Future work that performs more thorough analysis of the instrumental noise and data pipeline could help identify the source of this noise. For the purpose of this work, we simply fit for this excess uncorrelated noise, assuming it is Gaussian. Thus, the total error $\sigma_\textrm{tot}^2 = \sigma_\textrm{pipe}^2 + \sigma_\textrm{fit}^2$, where $\sigma_\textrm{pipe}$ is the nominal extraction error from our pipeline and $\sigma_\textrm{fit}$ is the error term we fit for. We used a separate $\sigma_\textrm{fit}$ for each spectral order, but assumed that $\sigma_\textrm{fit}$ is constant within an order. This seemed to be a suitable approximation based on our analysis of the residuals to the fit. 

In the analysis until now, we assumed that the width of the Gaussian trace of the fiber in the spatial dimension could be used as the LSF in the dispersion dimension. This assumption is not perfect as the NIRSPEC spectrograph was designed with a difference in focal lengths in the spatial and dispersion directions by a factor of 1.13 \citep{Robichaud1998}. In preliminary analysis of the OH sky lines which are unresolved at our resolution, we found that the LSF in the dispersion direction is indeed $1.12 \pm 0.02$ times wider than the width of the Gaussian profile measured in the spatial dimension. To conservatively account for any systematics in this preliminary measurement, we will allow the LSF width to vary between 1.0-1.2 times the width we measured in the spatial direction in Section \ref{sec:data-raw} (i.e., the aspect ratio of the 2D LSF). This corresponds to varying the resolution from $\sim$35,000 to $\sim$29,000. We note that since the stellar spectrum is built using empirical data, the LSF size only affects the broadening of the planetary model, $P_{LSF}$. 

We defined the log-likelihood to be:
\begin{equation}
    \ln(\mathcal{L}) = -\frac{1}{2} \sum \left( \frac{(\mathcal{H}[D_p] - \mathcal{H}[M_p])^2}{\sigma_{tot}^2} + \ln{(2\pi\sigma_{tot}^2)} \right),
\end{equation}
where we are summing over each spectral channel in the data. $\mathcal{H}[M_p]$ was constructed using Equation \ref{eq:hpmodel}. Note that implicit in $M_p$ are the parameters of the planet, the instrumental LSF, and the nuisance parameters of the stellar speckle spectrum. For each planet, we considered three different models: a forward model that only contains the stellar speckle spectrum and no planet signature (i.e., $\bar{\alpha}_p = 0$ and all planet parameters are fixed) which we call ``No Planet", a model containing a planet with no discernable rotation ($v\sin(i) = 0$ is fixed) which we call ``No Rotation", and a rotating planet model where all planet parameters are allowed to vary which we call ``Full Model". 

For HR 8799 b, c, and d, we used echelle orders 31 (2.44-2.49 $\upmu$m), 32 (2.36-2.41 $\upmu$m), and 33 (2.29-2.34 $\upmu$m) to fit our model to the data just as was done in Section \ref{sec:xcorr}. We did not use orders 34-39 (1.95-2.29 $\upmu$m) as they either had poor wavelength calibration (uncertainties larger than a spectral channel), or had strong CO$_2$ telluric absorption that was difficult to forward model accurately. For HR 8799 e, we only used orders 32 and 33 as we found that the strong telluric absorption features in order 31 could not be fully modeled, with correlated residuals of comparable amplitude to the planet signal. We omitted this order to mitigate the effect of residual telluric lines from biasing our fit of the planetary atmosphere. Future work to marginalize over localized telluric residuals \citep{Czekala2015} or improvements in modeling the stellar speckle spectrum could make this order useful in the fit, but we chose to omit this order for now. 

For the free parameters in each model, we used the following priors. For the planet properties, an uniform prior on $T_\textrm{eff}$ between 1000 and 1800 K, an uniform prior on $\log(g)$ between 3.5 and 5.5, an uniform prior on the planetary radial velocity between -150 and 150 km/s, an uniform prior on $v\sin(i)$ in the Full Model between 0 and 60 km/s, and an uniform prior on the planet flux from 0 to 25 DN (data numbers). For each order, we included a parameter for the stellar speckle flux and nuisance parameters to fit for systematics in the data: an uniform prior on the stellar speckles flux between 0 and 500 DN, an uniform prior on a residual background term between -10 and 10 DN due to imperfect background subtraction, and a log-uniform prior on the $\sigma_\textrm{fit}$ between 0.1 and 30 DN. In the Full Model and No Rotation models, we also included for each order a multiplicative factor to account for any broadening of the LSF beyond what we measured as the spatial width of the fiber trace with an uniform prior between 1.0 and 1.2. This term was not in the No Planet model because that model consisted solely of empirical data where we did not need to broaden anything to instrumental resolution. To fit the stellar speckle spectrum using a linear combination of on-axis stellar spectra, we used an uniform prior between 0 and 2 for each $c_i$ term. If we term $N_s$ as the number of stellar spectra, $D_{s,i}$, used to compute $D_s$ and $N_\textrm{order}$ as the number of spectral orders used in the fit, the No Planet model has $3N_\textrm{order} + N_s$ free parameters, the No Rotation model has $4 + 4N_\textrm{order} + N_s$ parameters, and the Full Model has $5 + 4N_\textrm{order} + N_s$ free parameters. In the end, this resulted in $\sim$20-30 free parameters, of which, we were mostly interested in the planetary parameters. 

\begin{figure*}
    \centering
    \includegraphics[width=0.75\textwidth]{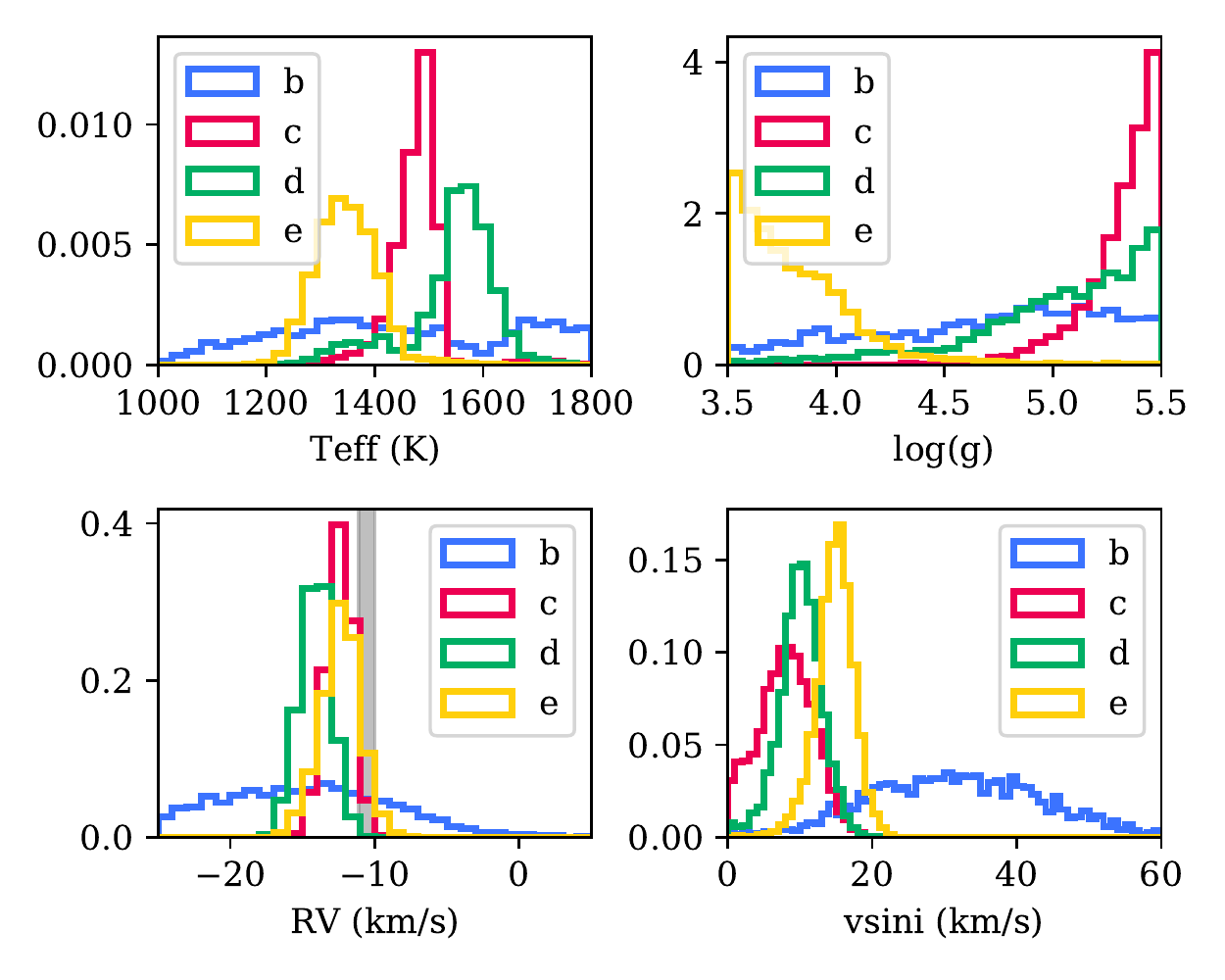}
    \caption{Posterior distributions of the main planetary parameters for each planet in the Full Model fit. In the bottom left panel, the gray shaded region represents the 68\% credible interval of the radial velocity of the host star relative to the Solar System barycenter. }
    \label{fig:posts}
\end{figure*}

We sampled the posterior using the nested sampling algorithm \citep{Skilling2004, Skilling2006} implemented in \texttt{pymultinest}, which allowed us to both perform parameter estimation and compute the model evidence \citep{Buchner2014}. The evidence, $P(D|M)$, is used to compute the Bayes factor, $B$, that can be used to assess the relative probability of model $M_1$ compared to $M_2$. If $P(M_1|D)/P(M_2|D)$ expresses the relative probability of the two models given the data, then,
\begin{equation}
    B \equiv \frac{P(D|M_1)}{P(D|M_2)} = \frac{P(M_1|D) P(M_2)}{P(M_2|D)P(M_1)},
\end{equation}
where $P(M)$ is the prior probability of that model. As we assumed each model has equal prior probability, then $B$ is equivalent to the relative probability of two models. We used the Bayes factor to compare the simpler models to the Full Model to determine whether the additional free parameters are justified. Thus, our $M_2$ will always be the Full Model. We listed the estimates on the planet parameters and the Bayes factor of each model compared to the Full Model of that planet in Table \ref{table:fit_params}. For the Full Model fits, we also plotted the posterior distributions of the key planet parameters in Figure \ref{fig:posts}. For the Full Model fits, the strongest covariance is between $T_\textrm{eff}$ and $\log(g)$ which we discuss in Section \ref{sec:atm}. We show corner plots of the posterior distributions from the Full Model fits in Appendix \ref{sec:corner}.

\begin{deluxetable*}{cc|ccccc|c}
\tablecaption{Model fits to HR 8799 KPIC data. For each parameter, the median value is listed, with the subscript and superscript values representing the range of the central 68\% credible interval with equal probability above and below the median (the central 95\% credible interval is listed in parentheses). \label{table:fit_params}}
\tablehead{
 Planet & Model & $T_\textrm{eff}$ (K) & $\log(g)$ & RV (km/s) & $v\sin(i)$ (km/s) &  Planet Flux (DN) & $B$
}
\startdata
b & Full Model & $1423.3^{+278.4(+362.3)}_{-212.6(-354.0)}$ & $4.8^{+0.4(+0.7)}_{-0.8(-1.2)}$ & $-15.6^{+6.5(+12.5)}_{-7.1(-14.9)}$ & $31.2^{+11.9(+22.5)}_{-11.7(-22.4)}$ & $5.4^{+1.9(+3.6)}_{-1.7(-3.0)}$ & 1.0 \\
b & No Rotation & $1448.5^{+275.4(+339.1)}_{-269.1(-399.4)}$ & $4.7^{+0.5(+0.8)}_{-0.7(-1.1)}$ & $-9.8^{+3.2(+6.6)}_{-5.9(-16.3)}$ & $-$ & $3.4^{+1.4(+2.8)}_{-1.1(-2.0)}$ & 0.13 \\
b & No Planet & $-$ & $-$ & $-$ & $-$ & $-$ & 0.092 \\
\hline
c & Full Model & $1482.6^{+26.2(+52.2)}_{-41.7(-118.0)}$ & $5.4^{+0.1(+0.1)}_{-0.2(-0.6)}$ & $-12.4^{+0.8(+1.6)}_{-1.1(-2.0)}$ & $8.1^{+3.8(+7.3)}_{-4.0(-7.3)}$ & $9.5^{+1.1(+4.0)}_{-0.8(-1.5)}$ & 1.0 \\
c & No Rotation & $1474.4^{+24.4(+38.9)}_{-36.3(-102.6)}$ & $5.4^{+0.1(+0.1)}_{-0.2(-0.4)}$ & $-12.4^{+0.9(+1.5)}_{-1.0(-1.9)}$ & $-$ & $9.0^{+0.7(+1.5)}_{-0.6(-1.3)}$ & 2.1 \\
c & No Planet & $-$ & $-$ & $-$ & $-$ & $-$ & $1.6 \times 10^{-41}$ \\
\hline
d & Full Model & $1558.8^{+50.9(+105.6)}_{-81.4(-238.7)}$ & $5.1^{+0.3(+0.4)}_{-0.4(-1.2)}$ & $-14.1^{+1.1(+2.2)}_{-1.2(-2.2)}$ & $10.1^{+2.8(+5.7)}_{-2.7(-6.6)}$ & $14.8^{+2.7(+4.8)}_{-3.8(-6.4)}$ & 1.0 \\
d & No Rotation & $1642.2^{+108.8(+150.5)}_{-127.1(-211.8)}$ & $5.4^{+0.1(+0.1)}_{-0.2(-0.6)}$ & $-15.2^{+1.4(+2.7)}_{-1.0(-1.8)}$ & $-$ & $11.2^{+1.8(+3.8)}_{-2.2(-4.1)}$ & 0.032 \\
d & No Planet & $-$ & $-$ & $-$ & $-$ & $-$ & $3.1 \times 10^{-29}$ \\
\hline
e & Full Model & $1345.6^{+57.0(+124.8)}_{-53.3(-99.5)}$ & $3.7^{+0.3(+0.9)}_{-0.1(-0.2)}$ & $-12.3^{+1.2(+2.5)}_{-1.4(-2.9)}$ & $15.0^{+2.3(+4.6)}_{-2.6(-6.1)}$ & $16.2^{+4.5(+7.7)}_{-3.6(-6.3)}$ & 1.0 \\
e & No Rotation & $1323.0^{+124.3(+318.3)}_{-87.6(-163.8)}$ & $4.3^{+0.7(+1.1)}_{-0.6(-0.8)}$ & $-14.6^{+1.1(+2.6)}_{-1.1(-1.9)}$ & $-$ & $9.2^{+1.7(+5.4)}_{-1.3(-2.3)}$ & 0.049 \\
e & No Planet & $-$ & $-$ & $-$ & $-$ & $-$ & $4.5 \times 10^{-20}$ \\
\enddata
\end{deluxetable*}

For each planet, we can ``decisively" reject any model with a $B$ that is more than 100 times smaller than the model with the highest $B$ based on the interpretation of the Bayes factor suggested by \citet{Jeffreys1983}. In Table \ref{table:fit_params}, we see that the No Planet model is ruled out for c, d, and e, but remains 9\% as likely as the Full Model for HR 8799 b. This finding is consistent with the marginal 3$\sigma$ detection of b using template cross correlation in Section \ref{sec:xcorr}, but we are now able to assign relative probabilities to the cases. The Bayes factor between a planet and no planet model offers an alternative way to determine detection significance rather than using CCF SNR, where the false alarm probability is unclear. Bayesian hypothesis testing methods have been used previously for source detection in cosmological datasets \citep{Carvalho2009} and for exoplanet direct imaging \citep{Golomb2019}. 

The data is less definitive in distinguishing between No Rotation and Full Model. In all cases, the No Rotation has a $>1$\% probability compared to the Full Model. We know that it is unphysical to assume these planets have no spin, but the No Rotation model is a good approximation for a planet with a spin that remains undetectable. Thus, we interpret this result as the current data does not provide a definitive detection for rotational broadening. This could be due a low signal to noise detection like in the case of HR 8799 b, but it could also be the difficulty of measuring small $v\sin(i)$ values, especially given the near-face-on orbital configuration of the system. Still, HR 8799 d and e have a No Rotation model with $B < 0.05$, which \citet{Jeffreys1983} interprets as ``very strong" evidence for rotational broadening: the Full Planet model being 30 times more likely for d and 20 times more likely of e. We note that $v\sin(i)$ values from the Full Model for both HR 8799 d and e are inconsistent with 0 by $> 3\sigma$, but the $B$ metric downweighs this to account for the addition of this free parameter that could cause overfitting. $B$ provides a more straightforward and more conservative assessment of the detection of rotational broadening rather than determining how far $v\sin(i)$ must deviate from 0 to be quantified as a detection, since the median $v\sin(i)$ will always be nonzero by construction. For HR 8799 b and c, we derive a 95\% upper limit on the $v\sin(i)$ of 51 and 14 km/s, respectively, based on their marginalized 1D posteriors plotted in Figure \ref{fig:posts}. Thus, in this work, we were able to make the first measurements or constraints of the rotation velocities of the HR 8799 planets. In the future, higher SNR detections or detections at higher spectral resolution would enable more definitive detections of rotation.

\begin{figure*}
    \centering
    \includegraphics[width=0.9\textwidth]{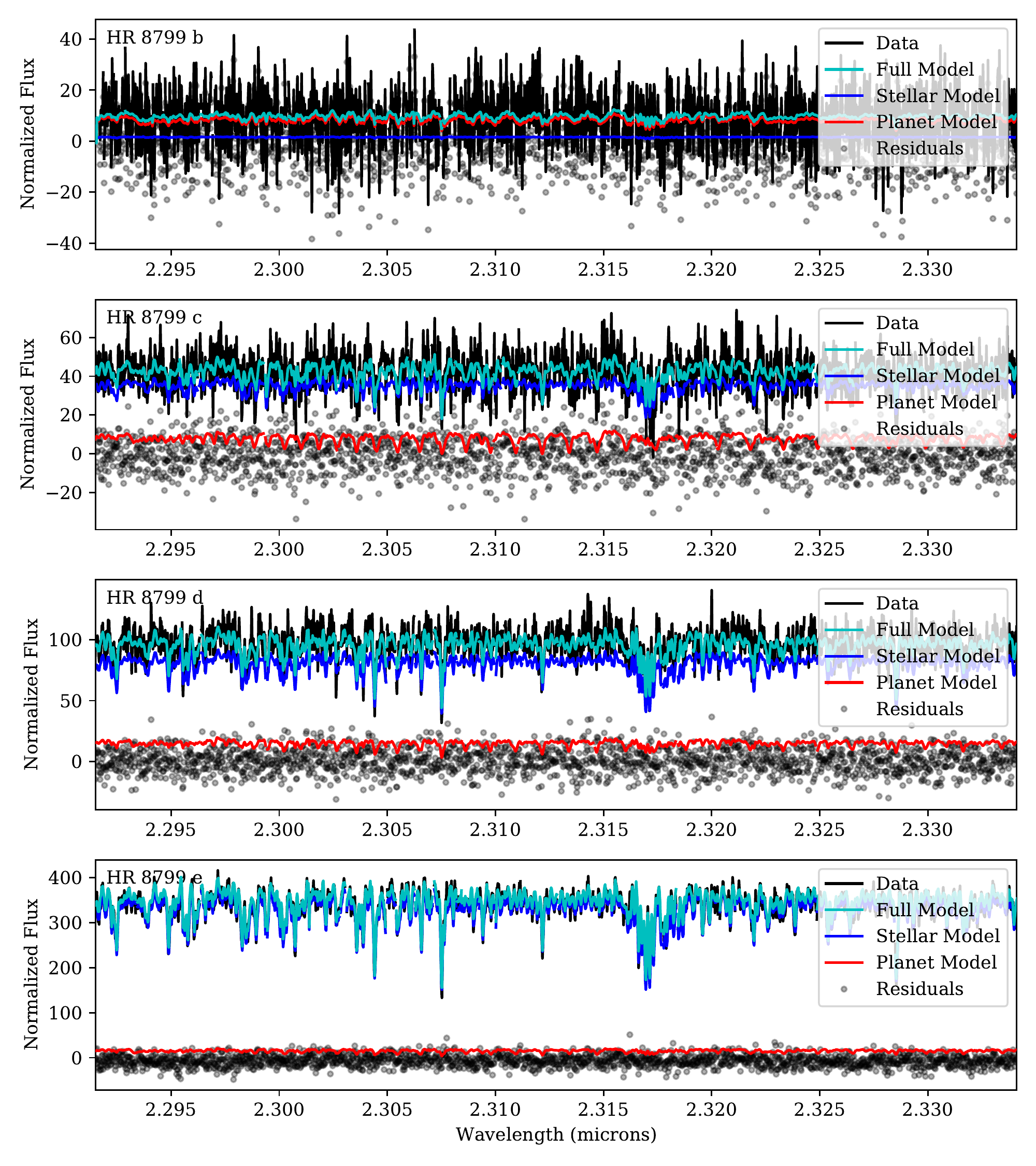}
    \caption{The 1D extracted spectra (black) from the fiber placed on each planet and the best fitting forward model from echelle order 33. The forward model (blue) has been deconstructed into its two constituent parts: the stellar model (cyan) built from a linear combination of on-axis stellar spectra and the planet model generated from the BT-Settl atmospheric models (red). The residuals to the fit are plotted as gray circles, and appear to be dominated by uncorrelated noise. A zoomed-in version of this plot is available in Appendix \ref{sec:res-acf}.}
    \label{fig:fm}
\end{figure*}

To assess the quality of our fits, we plotted the best fitting set of parameters from the Full Model fit to one order of the data for all four planets in Figure \ref{fig:fm}. Visual inspection indicated our forward models fit the data adequately, with the residuals appearing to be dominated by uncorrelated noise. We verified this by computing the autocorrelation function (ACF) of the residuals and found that the ACF is well approximated by a delta function, with the wings of the ACF having an amplitude of $\leq$5\% of the peak (see Appendix \ref{sec:res-acf}). This confirms that an uncorrelated noise model is sufficient, as we are likely dominated by thermal background noise of the instrument in our observations. 

\section{Discussion}\label{sec:discussion}

\subsection{Atmospheric Parameters}\label{sec:atm}
The detection of CO and \water{} but not \methane{} in our high-resolution spectra is consistent with previous atmospheric studies of the HR 8799 planets. Our CCF detections of HR 8799 c and d agree well with previous molecular cross-correlation analyses at lower resolution \citep[][Ruffio et al. in prep.]{Konopacky2013, WangJi2018}. Due to the weak detection of HR 8799 b in our limited observations, we were only able to marginally detect the combined signal of CO and \water{} and did not have enough SNR to address previous disagreements on the amount of methane in its atmosphere \citep{Barman2015, petit2018}. Longer integration times and improvements to instrument performance will improve on these data. Regardless, this is the first time HR 8799 e is studied at high spectral resolution, which demonstrates the high-contrast capabilities of KPIC. Previously, the highest resolution spectrum for HR 8799 e was the $R \sim 500$ GRAVITY spectrum \citep{GRAVITY2019}. For HR 8799 e, we found detections of CO and \water{} of similar strength as planets c and e and a similar non-detection of \methane{}. This is consistent with the picture that  the inner three planets have similar spectral signatures and luminosities based on lower resolution spectroscopy. 

\begin{figure*}
    \centering
    \includegraphics[width=0.35\textwidth]{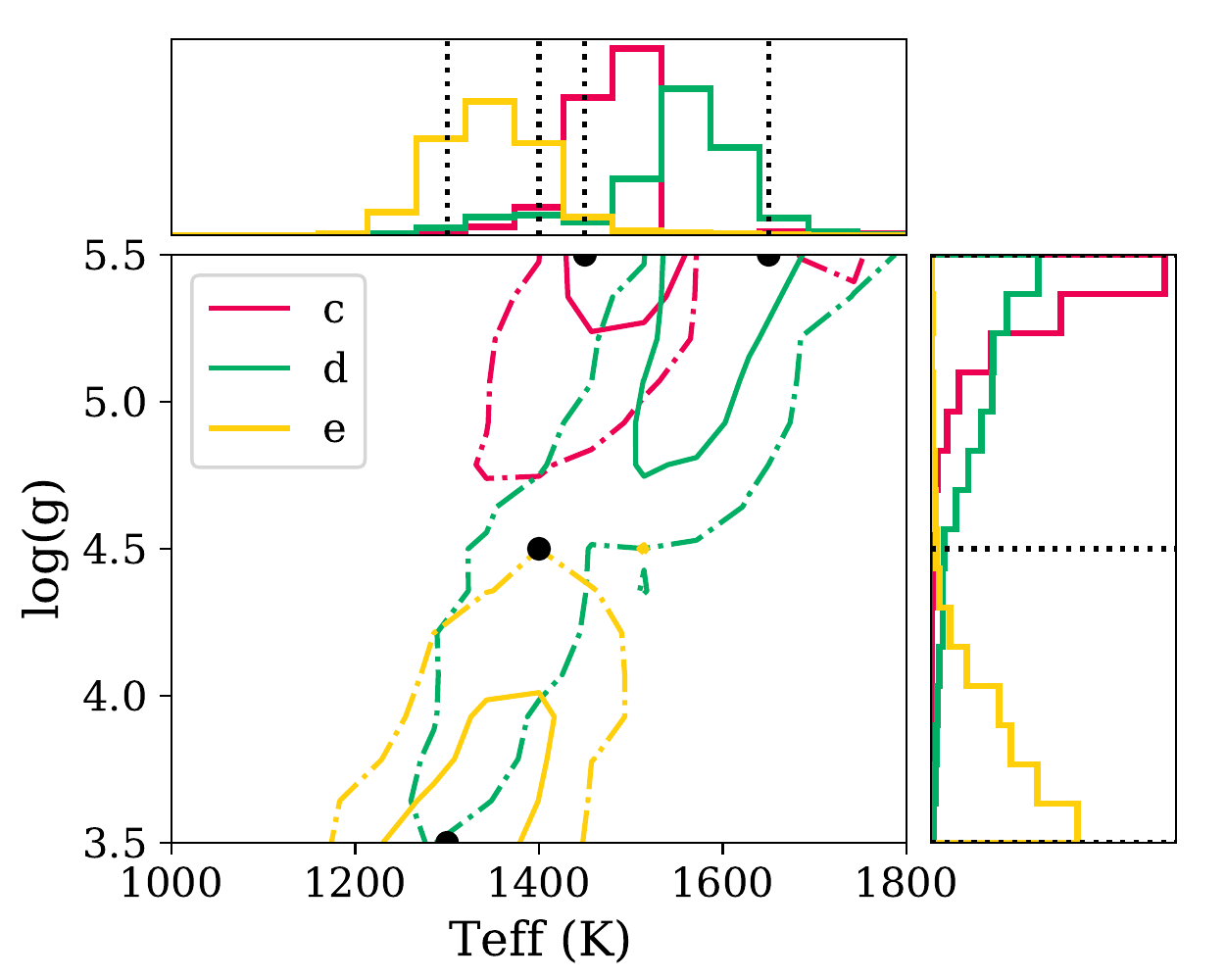}
    \includegraphics[width=0.60\textwidth]{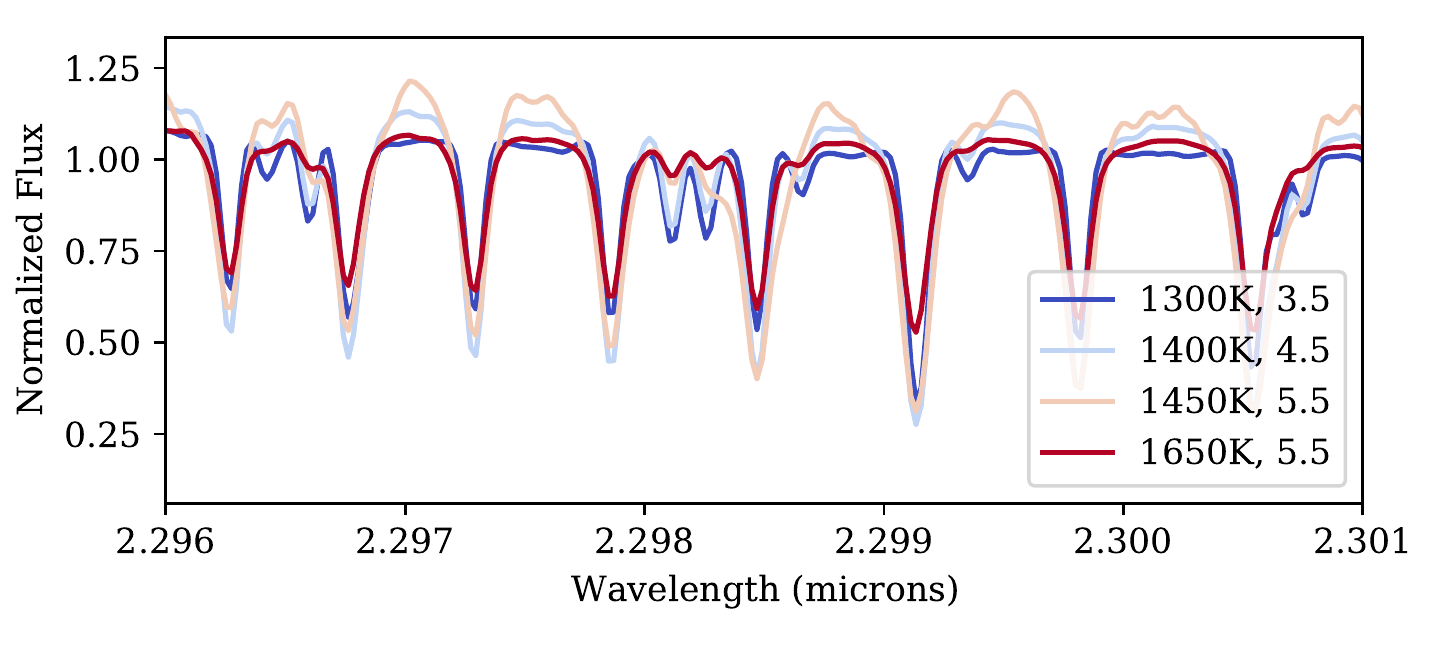}
    \caption{Atmospheric constraints on the HR 8799 planets. On the left are 1D and 2D marginalized posteriors of $T_\textrm{eff}$ and $\log(g)$ for HR 8799 c (red), d (green), and e (yellow). The 2D plot shows the 68\% (solid) 95\% (dot-dashed) credible region for each planet. A strong positive correlation between $T_\textrm{eff}$ and $\log(g)$ are observed. Black points and dotted lines correspond to representative models, which are plotted on the right.  }
    \label{fig:bts_vary}
\end{figure*}

We compared the planet fluxes measured in our forward model fits and reported in Table \ref{table:fit_params} to the expected fluxes of the planets using the end-to-end system throughputs reported in Table \ref{table:obs}, the exposure times of each frame, and the gain of the NIRSPEC detector \citep[3.03 e-/DN;][]{Lopez2020}. Using the photometry of the planets in the SPHERE $K2$-band from \citet{Zurlo2016}, we expected to measure a flux of 4, 13, 16, and 17 DN for HR 8799 b, c, d, and e respectively. These values are mostly consistent with our measured fluxes, although it appears that our inferred fluxes for HR 8799 c, d, and e to be lower on average. This could be due to losses in flux due to a misaligned fiber when the blind offset was performed. It is not entirely clear this is the case, since we do not see a similar effect for HR 8799 b, which had the largest blind offset. Still, the 95\% credible intervals of the fluxes are consistent with the literature photometry for all four planets. The fact the planet fluxes measured from our high resolution data agrees with broadband photometry increases our confidence that we have reliably separated the planet signal from the star. 

The bulk atmospheric properties (effective temperature and surface gravity) we obtained from our forward model fits somewhat agree with previous work at lower resolutions. Our $T_\textrm{eff}$ between 1400 and 1700~K for HR 8799 c and d agrees quite well with those obtained by \citet{bonnefoy2016} and \citet{Greenbaum2018} in their BT-Settl fits, but HR 8799 e with a $T_\textrm{eff}$ between 1200 and 1500~K is lower than those works by 300~K (only 0.02\% of our posterior have effective temperature $> 1650$~K). We note they obtained nonphysical radii in their BT-Settl fits which pinpoint to issues with fitting the BT-Settl grid to the HR 8799 spectral data (we did not perform absolute flux calibration so we did not constrain their radii). The $\log(g)$ values for HR 8799 e agree well with \citet{bonnefoy2016} and \citet{Greenbaum2018}. The posteriors for HR 8799 c and d favor $\log(g) > 4$ and peak at the upper bound of the prior (i.e., upper bound of model grid), which is at odds with those works which prefer $\log(g) < 4$ (only 0.05\% and 3\% of our posteriors for HR 8799 c and d, respectively, had $\log(g) < 4$). When looking beyond the BT-Settl model fits, our effective temperatures are systematically higher than the $\sim$1100~K derived from other atmospheric models or predicted by evolutionary models, and $\log(g)$ values for HR 8799 c and d continue to remain higher than literature values between 3.5 and 4 \citep{Marois2008, Marois2010, Konopacky2013, Currie2014, Ingraham2014, Skemer2014, bonnefoy2016, Greenbaum2018, Molliere2020}. This indicates there may be some systematic errors in the BT-Settl models, since both high and low resolution data both found higher $T_\textrm{eff}$ values for the BT-Settl model than all other models. One reason could be that BT-Settl models use equilibrium chemistry, and there has been evidence for disequilibrium chemistry in the HR 8799 planets causing stronger CO lines than expected and could make equilibrium chemistry models favor higher temperatures \citep{Skemer2014, Molliere2020}. However, we were not able to fit our data to other publicly available model grids with clouds as they are not generated at sufficiently high spectral resolution to match our data.

HR 8799 c, d, and e are thought to have similar bulk atmospheric properties (e.g., luminosity, effective temperature, radius) based on broadband spectroscopy and photometry \citep[e.g.,][]{Marois2010, bonnefoy2016}. This appears to be at odds with our marginalized 1D posteriors, which prefer different $T_\textrm{eff}$ and $\log(g)$ values for HR 8799 e. Partially, this is due to a strong correlation between $\log(g)$ and  $T_\textrm{eff}$: lower $T_\textrm{eff}$ values in our fits also prefer a lower $\log(g)$ (left panel of Figure \ref{fig:bts_vary}). The contours for all three planets lie on nearly the same plane, suggesting that the differences in both $T_\textrm{eff}$ and $\log(g)$ may stem from additional parameters that we were not able to adjust in the model grid fits (e.g., cloud properties or disequilibrium chemistry). The need to adjust more parameters has been seen in BT-Settl fits to substellar companions where additional extinction parameters have been used to augment the default cloud prescription \citep{Marocco2014, bonnefoy2016, Delorme2017, WardDuong2021}.

Looking more closely at why these fits to just our high-resolution spectra prefer these values, we plot how the BT-Settl spectra change as we change surface gravity and effective temperature in this correlated way in the right panel of Figure \ref{fig:bts_vary}. We observed that the spectra are quite similar visually. The $T_\textrm{eff} = 1300$~K, $\log(g)=3.5$ model preferred by HR 8799 e has similar CO line depths in many of the lines as the $T_\textrm{eff} = 1650$~K, $\log(g)=5.5$ model preferred by HR 8799 d, whereas the intermediate temperature models have the deepest CO lines. Since the BT-Settl models assumed solar abundances in carbon and oxygen, such changes in the \water{} and CO line depths could also be due to differences in abundance, given the broadband spectral data indicates the planets should have similar temperatures. Performing abundance measurements on this high resolution data in the future will offer independent and sensitive measurements of the C/O ratios of the planets. 

While we did not observe that planetary RV and $v\sin(i)$ are strongly correlated to $T_\textrm{eff}$ and $\log(g)$ (see Appendix \ref{sec:corner} for corner plots), any inconsistencies in these bulk atmospheric properties could bias our inferred planetary RV and $v\sin(i)$ values. This should partially be mitigated by the fits having broad priors in $T_\textrm{eff}$ and $\log(g)$, which allowed us to marginalize over a large range of possible models in deriving 1D posteriors for planetary RV and $v\sin(i)$. However, there could be remaining systematic errors in the model that are not accounted for. To examine if there could be additional biases, we fixed $T_\textrm{eff}$ and $\log(g)$ to the literature best-fit values for the BT-Settl grid from \citet{Greenbaum2018} for HR 8799 c, d, and e, and reran the fits. We found that planetary RV changed by $< 1\sigma$ and $v\sin(i)$ changed by $< 1.5\sigma$ for all three cases. We also assessed the sensitivity to cloud assumptions by rerunning the fits using the cloudless Sonora-Bobcat models instead, and found that both planetary RV and $v\sin(i)$ changed by $< 1\sigma$. These changes are comparable to our statistical errors, so any systematic biases should not significantly alter our results on RV and spin.  We note that the $B$ of these fits were $< 10^{-3}$ compared to the Full Model, so cloudless models are not good fits to the data, even if the planetary RV and $v\sin(i)$ appear to be consistent. Any smaller tweaks to the spectra (e.g., changing the C/O ratio) should have even smaller effects on the inferred planetary RV and $v\sin(i)$. 

\subsection{Orbital Velocity}
The planetary radial velocities we measured are not precise enough to add significant constraints on their orbits, but are consistent in amplitude and sign with previous work \citep{Wang2018, Ruffio2019}. Using the dynamically stable solutions from \citet{Wang2018} and the measurement that the position angle of the ascending node is less than 180 degrees \citep{Ruffio2019}, we expect an RV offset relative to the star of $1.9 \pm 0.2$~km/s for planet b, $0.0 \pm 0.2$~km/s for planet c, $-2.7 \pm 0.3$~km/s for planet d, and $-2.1 \pm 0.3$~km/s for planet e. We adopted the convention that negative RVs indicate the planet is moving towards us. The stellar radial velocity is somewhat uncertain with the latest estimate being $-10.5^{+0.5}_{-0.6}$~km/s with respect to the Solar System barycenter \citep{Ruffio2019}. 

Combining these two terms together, we have predicted planetary RVs relative to the Solar System barycenter of $-8.6^{+0.5}_{-0.6}$~km/s for planet b, $-10.5^{+0.5}_{-0.6}$~km/s for planet c, $-13.2^{+0.6}_{-0.7}$~km/s for planet d, $-12.6^{+0.6}_{-0.7}$~km/s for planet e. Although within the 95\% credible interval of all our measurements, our values for HR 8799 b, c, and d are systematically more negative by $\sim$1$\sigma$ ($\sim$1~km/s). This could indicate an error in the absolute wavelength solution of our instrument, or systematics in computing the stellar radial velocity in previous analyses at the 1~km/s level. Overall, these results confirm the ability to measure planetary radial velocities with KPIC.  


\subsection{Planetary Spin}

\begin{figure*}
    \centering
    \includegraphics[width=0.95\textwidth]{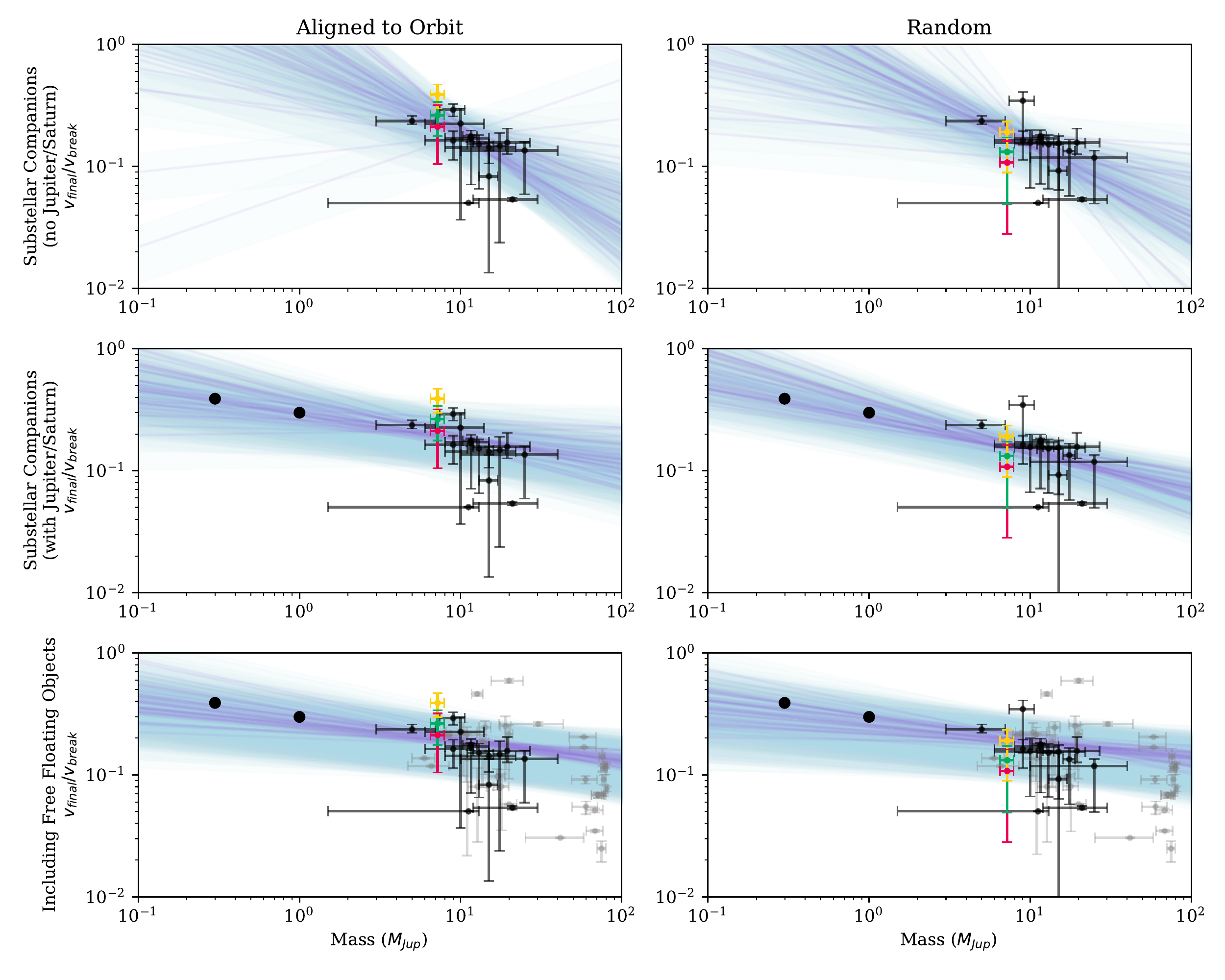}
    \caption{The final rotation rates relative to their breakup velocities as a function of mass for substellar objects. The left column assumes companion spins are aligned with their orbital planes, and the right column assumes randomly oriented spin axes. The top row only considers substellar companions excluding Jupiter and Saturn, the middle row includes Jupiter and Saturn, and the bottom row also includes free-floating substellar objects. In each plot, 50 possible power law relations using the fits from Table \ref{table:popfit} are plotted as purple lines with their dispersions shaded in blue. HR 8799 d is green and e is yellow. HR 8799 c is in red, but we note that one should consider this an upper limit. Other substellar companions with robust spin measurements are plotted in black. Free-floating objects are plotted in gray. Rotational velocities derived from photometrically measured rotation periods are unchanged between the two plots.  }
    \label{fig:pop_vel}
\end{figure*}

\begin{deluxetable*}{c|c|c|c|c|c|c|c|c|c}
\tablecaption{Derived rotation rate values for the HR 8799 planets. For non-detections of rotation, the 95\% upper limit or 5\% lower limit is reported. Otherwise, the values reported follow the convention in Table \ref{table:fit_params}. \label{table:rotvals}}
\tablehead{ 
 & &  \multicolumn{4}{c|}{Aligned to Orbit} & \multicolumn{4}{c}{Random} \\
Planet & $v_\textrm{break}$ & $v$ (km/s) & Period (h) & $v/v_\textrm{break}$ & $v_\textrm{final}/v_\textrm{break}$ & $v$ (km/s) & Period (h) & $v/v_\textrm{break}$ & $v_\textrm{final}/v_\textrm{break}$
}
\startdata 
b & $93^{+6(+13)}_{-6(-12)}$ & $< 131$ & $> 1.1$ & $ < 1.4$ & $ < 1.5$ & $< 108$ & $> 1.4$ & $ < 1.2$ & $ < 1.3$ \\
c & $103^{+6(+14)}_{-7(-13)}$ & $< 37$ & $> 4.0$ & $ < 0.36$ & $ < 0.39$ & $< 28$ & $> 5.4$ & $ < 0.27$ & $ < 0.31$ \\
d & $103^{+6(+14)}_{-7(-13)}$ & $25^{+8(+17)}_{-7(-16)}$ & $6.0^{+2.5(+11.3)}_{-1.5(-2.5)}$ & $0.24^{+0.08(+0.17)}_{-0.07(-0.16)}$ & $0.27^{+0.08(+0.18)}_{-0.08(-0.18)}$ & $13^{+7(+36)}_{-4(-8)}$ & $12.0^{+5.3(+22.6)}_{-4.4(-8.9)}$ &  $0.12^{+0.07(+0.34)}_{-0.04(-0.08)}$ & $0.13^{+0.08(+0.39)}_{-0.04(-0.09)}$ \\
e & $103^{+6(+14)}_{-7(-13)}$ & $37^{+8(+18)}_{-7(-15)}$ & $ 4.1^{+1.1(+3.3)}_{-0.8(-1.4)}$ & $0.36^{+0.08(+0.18)}_{-0.08(-0.16)}$ & $0.39^{+0.09(+0.19)}_{-0.08(-0.17)}$ & $18^{+10(+47)}_{-4(-8)}$ & $ 8.4^{+2.3(+6.4)}_{-3.0(-6.1)}$ & $0.17^{+0.10(+0.47)}_{-0.04(-0.07)}$ & $0.19^{+0.10(+0.53)}_{-0.04(-0.08)}$ 
\enddata
\end{deluxetable*}

\begin{deluxetable*}{c|c|c|c}
\tablecaption{Power law fits to the rotation rate of gas giant planets. The values reported follow the convention in Table \ref{table:fit_params}. \label{table:popfit}}
\tablehead{ 
Case & $C$ & $\beta$ & $\sigma_f$ 
}
\startdata 
Prior & LogUniform(0.01, 0.9) & Uniform(-3, 3) & LogUniform(0.01, 0.5) \\
\hline
Aligned to Orbit & \multicolumn{3}{c}{} \\
\hline
Substellar Companions (no Jupiter/Saturn) &  $0.20^{+0.03(+0.06)}_{-0.02(-0.04)}$ & $-0.56^{+0.32(+0.66)}_{-0.09(-0.21)}$ & $0.39^{+0.07(+0.10)}_{-0.08(-0.14)}$ \\
Substellar Companions (with Jupiter/Saturn) & $0.20^{+0.03(+0.07)}_{-0.02(-0.04)}$ & $-0.18^{+0.10(+0.18)}_{-0.11(-0.24)}$  & $0.42^{+0.05(+0.08)}_{-0.07(-0.14)}$ \\
Including Free Floating Objects & $0.19^{+0.02(+0.03)}_{-0.01(-0.03)}$ & $-0.15^{+0.05(+0.09)}_{-0.05(-0.11)}$ & $0.47^{+0.02(+0.03)}_{-0.03(-0.07)}$ \\
\hline
Random & \multicolumn{3}{c}{} \\
\hline
Substellar Companions (no Jupiter/Saturn) & $0.15^{+0.02(+0.04)}_{-0.02(-0.03)}$ & $-0.61^{+0.25(+0.53)}_{-0.27(-0.57)}$  & $0.33^{+0.09(+0.16)}_{-0.07(-0.12)}$ \\
Substellar Companions (with Jupiter/Saturn) & $0.15^{+0.01(+0.03)}_{-0.01(-0.03)}$ & $-0.30^{+0.07(+0.14)}_{-0.09(-0.21)}$ & $0.33^{+0.08(+0.15)}_{-0.06(-0.11)}$ \\
Including Free Floating Objects & $0.19^{+0.02(+0.03)}_{-0.01(-0.02)}$ & $-0.14^{+0.05(+0.09)}_{-0.05(-0.10)}$ & $0.47^{+0.02(+0.03)}_{-0.03(-0.07)}$ \\
\enddata
\end{deluxetable*}

We used our measured $v\sin(i)$ values to compare the rotation velocities for HR 8799 c, d, and e to the population of substellar companions and free-floating objects with measured spins. Although the spin measurements for HR 8799 c and d are not definitive, the strong evidence for rotation still gave us confidence to assess what these measurements would imply about their rotational evolution. We excluded HR 8799 b in the analysis since its $v\sin(i)$ is unconstrained given the tentative detection, and any preference in $v\sin(i)$ values may be spurious (we still reported some numbers taking the posterior at face value). 

The bulk of the companion population was studied in \citet{Bryan2020}. We also did not consider HD 106906 b and AB Pic b in our subsequent analyses, since they only have tenuous rotation measurements \citep{Zhou2019, Zhou2020}. We included the $v\sin(i)$ measured for GQ Lup b \citep{Schwarz2016}, the rotation period measured for HN Peg b \citep{Zhou2016}, and the rotation period of Jupiter \citep{Dessler1983} and Saturn \citep{Helled2015} to extend the mass range of substellar companions considered. For the free-floating objects, we included the planetary mass objects from \citet{Bryan2020} as well as variable brown dwarfs with photometrically measured rotation periods compiled in \citet{Vos2020} and supplemented with measurements from \citet{Tannock2021}. For the variable brown dwarfs, we only used those with parallax measurements from the literature \citep{Faherty2012,Smart2013,Tinney2014,liu2016,Theissen2018,Gaia2018,Best2020,GaiaEDR3} in order to convert their $K$-band fluxes \citep{Cutri2003} to masses using evolutionary tracks \citep{Baraffe2003}. We note that this sample contains objects formed through a variety of formation mechanisms: free-floating objects likely are the low-mass tail of cloud fragmentation \citep[e.g.,][]{Luhman2012}, while planets like Jupiter and Saturn formed via core-accretion \citep{Pollack1996}. Thus, depending on the dominant processes that produced the spins we measured, these populations may or may not obey similar spin trends. However, \citet{Bryan2020} recently argued for a single spin regulation mechanism for both free-floating objects and companions.

Converting $v\sin(i)$ measurements requires accounting for the inclination degeneracy, and for most of these objects, including the HR 8799 planets, there are no empirical constraints on the spin axis inclination, and thus obliquity. While \citet{Vos2020} found a correlation between near-infrared color anomaly with spin axis inclination for low-gravity brown dwarfs, this relation is not particularly constraining for the HR 8799 planets: they have $(J-K)\sim2.5$~mag \citep{Zurlo2016} that corresponds to a $(J-K)$ color anomaly of $\sim-0.1$~mag \citep{liu2016}, which lands in a region where inclinations between $20^\circ$ and $90^\circ$ were measured \citep{Vos2020}. Directly measuring the inclination of the spin axis typically requires a photometrically derived rotation period and a rotational broadening measurement, and an obliquity constraint needs the spin axis inclination and the orientation of the orbital plane. The planetary-mass companion 2M0122b has the only exoplanetary obliquity measurement to date and its obliquity may be high \citep{Bryan2020a}. A scenario that can produce this companion and give it a misaligned obliquity is formation via gravitational instability, where gravito-turbulence in the disk can torque fragment spin axes out of alignment \citep{Bryan2020a}. For the HR 8799 planets, formation via core accretion is preferred based on occurrence rate statistics and atmospheric compositions \citep{Konopacky2013, Barman2015, Nielsen2019, Molliere2020}. In this case, obliquity excitation may arise from chaotic spin-orbit dynamics due to the near commensurability of nodal and spin precession frequencies \citep[e.g.][]{NerondeSurgyLaskar1997,LiBatygin2014,ShanLi2018,Saillenfest+2019}.  Specifically, consider the planet $i = b,c,d,e$, with mass $m_i$, radius $R_i$, spin frequency $\Omega_i = \bar \Omega_i/(G m_i/R_i^3)^{1/2}$, and semi-major axis $a_i$, orbiting a host star of mass $M_\star$.  Because the nodal precession rate of planet $i$ due to the other planets is of order \citep[e.g.][]{PuLai2018}
\begin{equation}
\omega_i \approx \sum_{j\ne i} \omega_{ij},
\hspace{5mm}
\omega_{ij} \approx \frac{3 G m_j a_<^2}{4 a_>^3 \sqrt{G M_\star a_i}},
\end{equation}
where $a_< = \min(a_i,a_j)$ and $a_> = \max(a_i,a_j)$, while the spin-precession rate of planet $i$ due to the host star's tidal torque is (assuming a fully convective planet, e.g. \citealt{Lai2014})
\begin{equation}
\alpha_i \approx \frac{3 G M_\star}{4 a_i^3} \left( \frac{R_i^3}{G m_i} \right)^{1/2} \bar \Omega_i,
\end{equation}
and all spin frequencies have magnitudes $\bar \Omega_i \sim 0.3$, we have for the HR 8799 planets, $\omega_b/\alpha_b \sim 8.1$, $\omega_c/\alpha_c \sim 5.9$, $\omega_d/\alpha_d \sim 2.8$, and $\omega_e/\alpha_e \sim 0.6$, so all ratios of $\omega_i/\alpha_i$ are of order unity.  Hence, this system may be susceptible to secular spin-orbit resonances \citep[e.g.][]{MillhollandLaughlin2019, Bryan2020}.  The coupled inclination and obliquity dynamics of the HR 8799 system will be investigated in a future work.

In the following analysis, we considered two bounding assumptions: the spin axes being randomly oriented in space as was done in \citet{Bryan2020}, and spin axes being aligned with the orbital planes based on orbital inclination posteriors derived from the literature \citep{Ginski2014, Bryan2016, Wang2018, Pearce2019, Bryan2020a, Bowler2020, Nowak2020b}. We considered these to be bounding assumptions as reality likely does not have perfectly aligned obliquities nor a large fraction of retrograde obliquities. Two companions (SR 12 c and 2M0249b) did not have measurements of orbital motion or photometrically derived rotation periods, so the priors on their spin axes were assumed to be isotropic in both cases. Similarly, the free-floating objects were assumed to have isotropic spin axes orientations. 

In both cases, we computed the rotational velocities ($v$), with a preference of using photometrically derived rotation periods over $v\sin(i)$ measurements, since typically radius uncertainties are smaller than the inclination uncertainties. For the HR 8799 planets, we used an orbital inclination of $i = 24^\circ \pm 3^\circ$ for the zero obliquity case \citep{Wang2018}. We compared the rotation velocities against their break-up velocities, which we define simply as:
\begin{equation}
    v_\textrm{break} = \sqrt{GM/R},
\end{equation}
where $M$ is the mass of a object and $R$ is its radius. This ignores the effect of oblateness, which will decrease the break-up velocity, with the exact decrease depending on the moment of inertia of the bodies \citep{Marley2011,Tannock2021}. To derive breakup velocities for the HR 8799 planets, we used masses of $7.2 \pm 0.7$~$M_\textrm{Jup}$ for the inner three planets, a mass of $5.8 \pm 0.5 $~$M_\textrm{Jup}$ for HR 8799 b, and radii of $1.2 \pm 0.1$~$R_\textrm{Jup}$ \citep{Marois2008, Marois2010, Wang2018}. The values of $v_\textrm{break}$ for each planet are listed in Table \ref{table:rotvals}. Under both obliquity assumptions, we computed $v$, rotation period, and $v/v_\textrm{break}$, and listed the values for the HR 8799 planets in Table \ref{table:rotvals}. Note that we only reported 95\% upper limits for HR 8799 b and c. Although no rotation periods have been measured from photometric monitoring of the HR 8799 planets \citep{Apai2016,Biller2021}, our inferred rotation periods are $> 3$~hours for HR 8799 d and e under both obliquity assumptions. This is consistent with the rotation periods of PSO J318.5-22 and 2M1207b, the only objects similar in mass and age with photometrically measured rotation periods \citep{Zhou2016,Biller2018}. The rotation periods are also consistent with the picture that predicted magnetic braking from the circumplanetary disk regulating the spin of the planet at very early times ($\lesssim 1$~Myr): \citet{Bryan2018, Bryan2020} found spins between 5-20\% of breakup for planetary mass companions, \citet{Batygin2018} predicted a terminal rotation period prediction of $\sim$9 hours, and \citet{Ginzburg2020} placed a maximum rotation period of 29-43 hours.

To compare these rotation rates against the whole population of substellar objects with measured rotation rates, we needed to account for the fact the rotation speed of an object depends on age. In our sample, ages ranged from $10^6$ to $10^9$ years old. In the absence of outside influence, an isolated body will spin up as its radius contracts due to radiative cooling with $v \propto R^{-1}$ and $v/v_\textrm{break} \propto R^{-1/2}$. The current population of planetary mass objects has been shown to be consistent with this trend \citep{Bryan2020}. We evolved the radii of all companions from their current radii to the radii predicted by hot-start evolutionary models at 5 Gyr \citep{Baraffe2003} and computed their final spin velocities relative to their final break-up velocities, $v_\textrm{final}/v_\textrm{break}$. The values of $v_\textrm{final}/v_\textrm{break}$ for the HR 8799 planets are listed in Table \ref{table:rotvals}.

For the youngest objects in the sample, they may still host a circumplanetary disk (CPD) which combined with planetary magnetic fields is thought to regulate planetary spin \citep{Batygin2018, Ginzburg2020}. CPD lifetimes are poorly constrained, so we do not know for sure which objects are currently experiencing magnetic breaking. With evidence that CPDs and accretion occurs for the PDS 70 planets \citep{Haffert2019,Isella2019}, which are $8 \pm 1$~Myr \citep{Wang2020}, we assumed CPDs and regulation of planetary spin by magnetic braking occurs up to 10 Myr. For companions younger than 10~Myr, we made one modification and assumed their rotational velocities are constant until 10~Myr. We note that magnetic braking is actually spinning down the planets, although the time dependence is weak (see Figure 1 of \citealt{Ginzburg2020}), so this is a rough approximation.  

We compared the $v_\textrm{final}/v_\textrm{break}$ values of the HR 8799 planets against the rest the gas giant and brown dwarf population in Figure \ref{fig:pop_vel} under our bounding obliquity assumptions and with three different subsets of the data (substellar companions excluding Jupiter and Saturn, substellar companions including Jupiter and Saturn, and all objects including free floating substellar objects). 
Regardless of obliquity assumption and data subset, Figure \ref{fig:pop_vel} appears to show a trend of increasing rotational velocity (relative to break-up velocity) with decreasing companion mass. A similar trend in $v$ and mass for substellar objects has been pointed out in previous works \citep{Snellen2014, Scholz2018}, but no significant trend was found when looking at 1-20 $M_\textrm{Jup}$ planetary mass objects previously \citep{Bryan2018, Zhou2019, Xuan2020}. \citet{Batygin2018} and \citet{Ginzburg2020} suggested that lower mass planets may be less effective at ionizing their CPDs, making magnetic braking less effective in spinning down the planets and creating a spin speed that is mass dependent. To quantify the influence of mass on rotation rates, we followed the methodology from \citet{Wolfgang2016} where they used hierarchical Bayesian modeling to derive a probabilistic mass-radius relationship for transiting exoplanets. This approach accounts for uncertainties in both the independent and dependent variables by using their posteriors in the fit. This is important for our study as the constraints on mass and rotation speed can be weak and non-Gaussian, so point-estimates such as the median and 68\% credible interval do not provide the full picture. We fit a model that is a power-law with intrinsic scatter of the following functional form:
\begin{equation}
    \frac{v_\textrm{final}}{v_\textrm{break}} \sim \textrm{Normal}\left(\mu = C\left(\frac{M}{10 M_\textrm{Jup}}\right)^\beta, \sigma=\sigma_f \mu \right).
\end{equation}
Following the notation of \citet{Wolfgang2016}, the $\sim$ above indicates that the rotation rate is ``drawn from the Normal distribution" with a mean ($\mu$) and standard deviation ($\sigma$) that varies with companion mass ($M$). We fit for the power law constant ($C$), the power law index ($\beta$), and the fractional dispersion of the distribution at a given mass ($\sigma_f$). For both spin axis inclination assumptions and for the three different data subsets, we estimated these population-level parameters and marginalized over the priors on rotation velocities and masses for the individual companions using the \texttt{dynesty} nested sampling package \citep{Speagle2020} with multiple bounding ellipsoids \citep{Feroz2009} and random slice sampling \citep{Neal2003,Handley2015a,Handley2015b}. The priors and inferred values of the population-level parameters are listed in Table \ref{table:popfit} and the fits excluding the Solar System gas giants are plotted in Figure \ref{fig:pop_vel}.

In all cases, we found a negative power law index was preferred. With the exception of the case where we looked at the substellar companions excluding Jupiter and Saturn and assuming zero obliquity, the 95\% credible interval of $\beta$ was entirely negative, although the mass dependence is weak: the 95\% credible interval for $\beta$ is between -0.5 and 0 when including the Solar System gas giants, and between -1.2 and +0.1 when excluding them. This could be a tentative sign that the degree of ionization of the CPD by the planet itself affects the final rotation rate of a planet. However, there are other ways to create a mass dependence in $v/v_\textrm{break}$: for example, the effect of oblateness would cause $v_\textrm{break}$ to decrease by a factor that depends on the object's moment of inertia, which itself is also mass dependent \citep{Marley2011,Tannock2021}. Even if it is true, the large scatter in the relation ($\sigma_f$ ranging from 25-50\%) indicates there would still be other variables at play. For example, since the planet is free to spin-up again after the CPD disperses in the magnetic braking scenario, different CPD lifetimes could also contribute to the scatter in the population \citep{Bryan2020}. Such a scatter is seen in young free-floating brown dwarfs where a rotation period dispersion was linked to the presence or absence of a disk at $\sim$8~Myr ages \citep{Scholz2018,Moore2019}. In our fits, including the free-floating objects caused $\sigma_f$ to reach the upper bound of the prior (50\% dispersion), which may indicate that mass is not a driving factor in shaping the spins of the free-floating objects, or that they are affected in a different way, since they formed through a different process than planetary companions. As the spins of more companions in the 1-10 $M_\textrm{Jup}$ range are measured with instruments like KPIC, REACH, and HiRISE, we will be able to more confidently assess whether there is a trend in their rotation velocity with mass, and if there are other factors at play. 

\section{Conclusions}\label{sec:conc}

We obtained high spectral resolution ($R \sim 35,000$) $K$-band spectra of all four HR 8799 planets taken as part of the science-demonstration of KPIC, a new instrument that combines high contrast imaging with high resolution spectroscopy to enable HDC techniques. By cross-correlating the spectra with molecular templates, we detected both CO and \water{} at high significance ($> 10\sigma$ combined) for HR 8799 c, d, and e, and tentatively detected HR 8799 b at $3\sigma$. These are the first detections of HR 8799 b, d, and e at high spectral resolution ($R \gtrsim$ 10,000), where we can fully resolve the individual molecular lines in the planets' atmospheres. The detection of HR 8799 c has an SNR that is 2$\times$ better while using 3.5$\times$ less exposure time than the previous NIRSPAO detection \citep{WangJi2018}. With KPIC, we are able to access closer in planets such as HR 8799 e (385~mas from its star), which has previously never been detected at either medium or high resolution, demonstrating the HDC capabilities of KPIC. These are the most challenging directly imaged exoplanets characterized with high resolution spectroscopy to date given their flux ratios and angular separations: while $\beta$ Pic b \citep{Snellen2014} is slightly closer in than HR 8799 e, it is an order-of-magnitude brighter in the $K$ band, which makes it easier to isolate from the diffracted stellar light.

Beyond molecular detections via cross correlation, we developed a likelihood-based approach to jointly fit the stellar speckles and planet light in our high resolution data. We used the Bayes factor computed from our framework to assess detection probabilities that corroborated the molecular CCF analysis: definitive detections of HR 8799 c, d, and e, but a tentative detection of HR 8799 b where the No Planet model is still 9\% as likely as the best model fit with a planet. Using our high resolution data alone, we inferred bulk properties from their atmospheres using the BT-Settl atmospheric models \citep{Allard2012}. The measured radial velocities of the four planets are consistent with previous orbital constraints \citep{Wang2018, Ruffio2019}. We found HR 8799 c, d, and e to have $T_\textrm{eff}$  constrained to between 1200 and 1600~K, which is somewhat consistent with previous works that used the BT-Settl models \citep{bonnefoy2016,Greenbaum2018}, but significantly higher than other literature fits that used different model grids. We also favored values of $\log(g)$ above 4.5 for HR 8799 c and d, which are inconsistent with fits at low spectral resolution and predictions of mass and radius from evolutionary models. We speculated that deficiencies in the BT-Settl models could be partially responsible for our inferred parameters. Future work to study these planets with more accurate atmospheric models would be able to determine whether our results are due to model or data systematics, but we were not able to perform such an analysis due to the limited number of publicly available forward model grids that both model clouds and support a spectral resolution of 35,000.

We found strong evidence for rotational broadening in our spectra of HR 8799 d and e, although more data is needed to make them decisive detections. Taken at face value, our inferred $v\sin(i) = 10.1^{+2.8}_{-2.7}$~km/s for HR 8799 d and $v\sin(i) = 15.0^{+2.3}_{-2.6}$~km/s for HR 8799 e correspond to $0.24^{+0.08}_{-0.07} v_\textrm{break}$ and $0.36^{+0.08}_{-0.08} v_\textrm{break}$ assuming zero obliquities and $0.12^{+0.07}_{-0.04} v_\textrm{break}$ and $0.17^{+0.10}_{-0.04} v_\textrm{break}$ assuming random oblitquities. We found a rotational velocity upper limit for HR 8799 c of $< 14$~km/s corresponding to $< 0.36 v_\textrm{break}$ in the zero obliquity case, but the spin of HR 8799 b was essentially unconstrained given the tentative detection. These spin measurements are consistent with the picture of magnetic braking regulating the spins of gas giant planets during the planet formation process. When combining these rotation measurements with literature rotation measurements of substellar companions as well as our Solar System gas giants, we found a tentative trend of increasing rotation rate with decreasing planet mass that could point to magnetic braking being less efficient for lower mass planets. 

\subsection{Future Prospects}

More spin measurements of 1-10~$M_\textrm{Jup}$ companions will test whether this mass-spin relation holds, or if there are other key parameters to control the spin rate of giant planets. In our analysis, we assumed two bounding assumptions on the obliquity of the planets to derive true rotation rates. Obliquity constraints on the HR 8799 planets themselves may be possible in the future if their periods can be derived from rotational modulations in their light curves. This could be possible with new hardware designed to improve the photometric calibration of coronagraphic systems \citep{Bos2020}. 

Future analysis work that moves beyond molecular detection and characterization of bulk parameters will provide more insights into the HR 8799 planets. Retrievals of the chemical abundances of their atmospheres have been applied to broadband photometry and lower resolution spectra \citep{Lavie2017,Molliere2020}. The likelihood framework we constructed to fit the high spectral resolution KPIC data would be straightforward to include in these Bayesian retrievals, allowing us to make use of the information at all spectral resolutions in understanding the chemical makeup of these planets. KPIC is also being used to conduct a spectroscopic survey of substellar companions which were previously too close to their bright host stars to study at high spectral resolution with slit spectrographs. Measuring planet characteristics like spin, orbital parameters, and composition across this population will allow us to look for trends that would pinpoint the dominant processes in their formation and evolution. 

Lastly, future planned upgrades to KPIC \citep{Pezzato2019,Jovanovic2020} are designed to improve the coupling of planet light, reduce the coupling of star light, and mitigate the instrument thermal background. As we are limited by thermal background noise currently, increasing the amount of planet light reaching the detector and reducing the amount of thermal background photons will directly translate into improvements in the SNR of KPIC observations. Re-observing the HR 8799 planets with these upcoming upgrades would give us more precise planetary spin measurements and improved sensitivity to any trace methane abundances in their atmospheres to constrain the level of disequilibrium chemistry. The initial phase of KPIC is the bare-bones hardware necessary to enable HDC techniques, and refined HDC techniques will allow for detailed characterization of all of the directly imaged exoplanets.  

\begin{acknowledgments}
We thank Sivan Ginzburg, Konstantin Batygin, and Eve Lee for useful discussions. We thank the Keck staff for helping support the deployment of KPIC during the global pandemic. 
J.J.W. is supported by the Heising-Simons Foundation 51 Pegasi b Fellowship. This work was supported by the Heising-Simons Foundation through grants \#2015-129, \#2017-318 and \#2019-1312. This work was supported by the Simons Foundation. 
ECM is supported by an NSF Astronomy and Astrophysics Postdoctoral Fellowship under award AST-1801978.
Part of the computations presented here were conducted on the Caltech High Performance Cluster, partially supported by a grant from the Gordon and Betty Moore Foundation. Part of this work was carried out at the Jet Propulsion Laboratory, California Institute of Technology, under contract with NASA. 
Data presented in this work were obtained at the W. M. Keck Observatory, which is operated as a scientific partnership among the California Institute of Technology, the University of California and the National Aeronautics and Space Administration. The Observatory was made possible by the generous financial support of the W. M. Keck Foundation. We wish to recognize and acknowledge the very significant cultural role and reverence that the summit of Maunakea has always had within the indigenous Hawaiian community. We are most fortunate to have the opportunity to conduct observations from this mountain. 
\end{acknowledgments}

\facility{Keck (KPIC)}

\software{\texttt{astropy} \citep{Astropy2018}, \texttt{scipy} \citep{2020SciPy-NMeth}, \texttt{PyAstronomy} \citep{pyastronomy}, \texttt{pymultinest} \citep{Buchner2014}, \texttt{dynesty} \citep{Speagle2020}, \texttt{whereistheplanet} \citep{whereistheplanet}, \texttt{corner} \citep{corner} }

\bibliography{kpic_paper}

\begin{appendix}

\section{Corner plots for Full Model Fits}\label{sec:corner}
\begin{figure}
    \centering
    \plotone{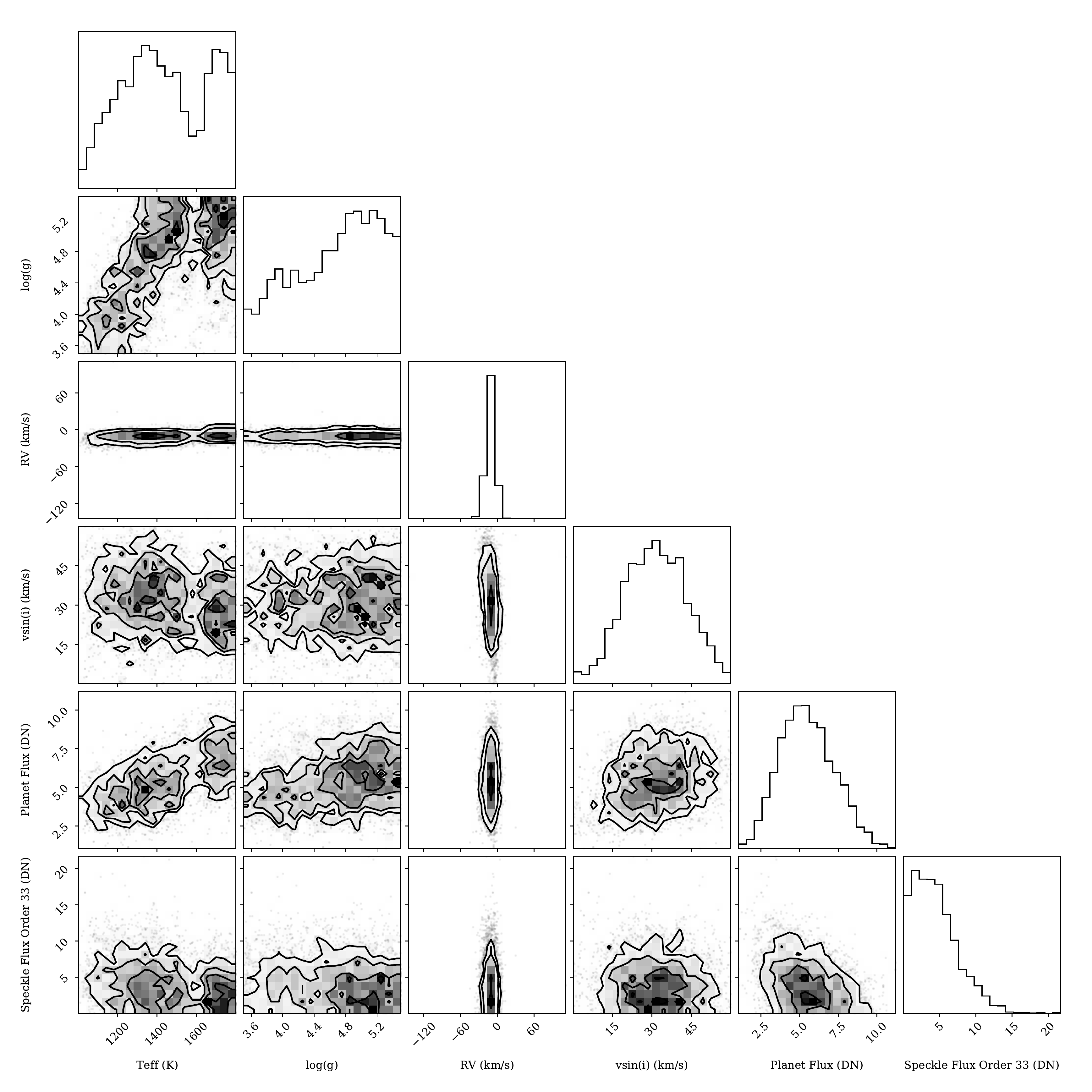}
    \caption{Corner plot for HR 8799 b of the posterior distribution of key parameters in the Full Model fit. \label{fig:corner-b}}
\end{figure}

\begin{figure}
    \centering
    \plotone{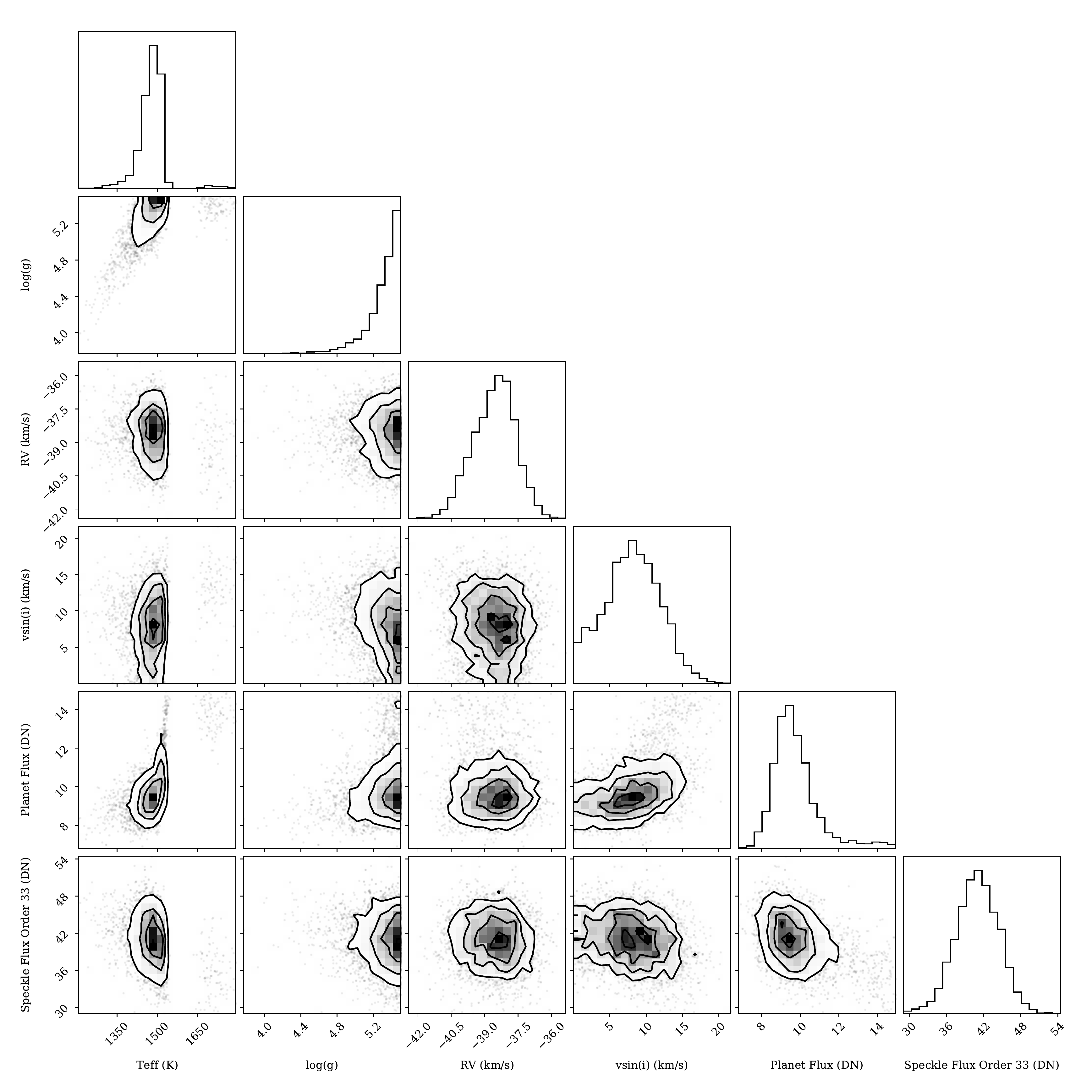}
    \caption{Corner plot for HR 8799 c of the posterior distribution of key parameters in the Full Model fit. \label{fig:corner-c}}
\end{figure}

\begin{figure}
    \centering
    \plotone{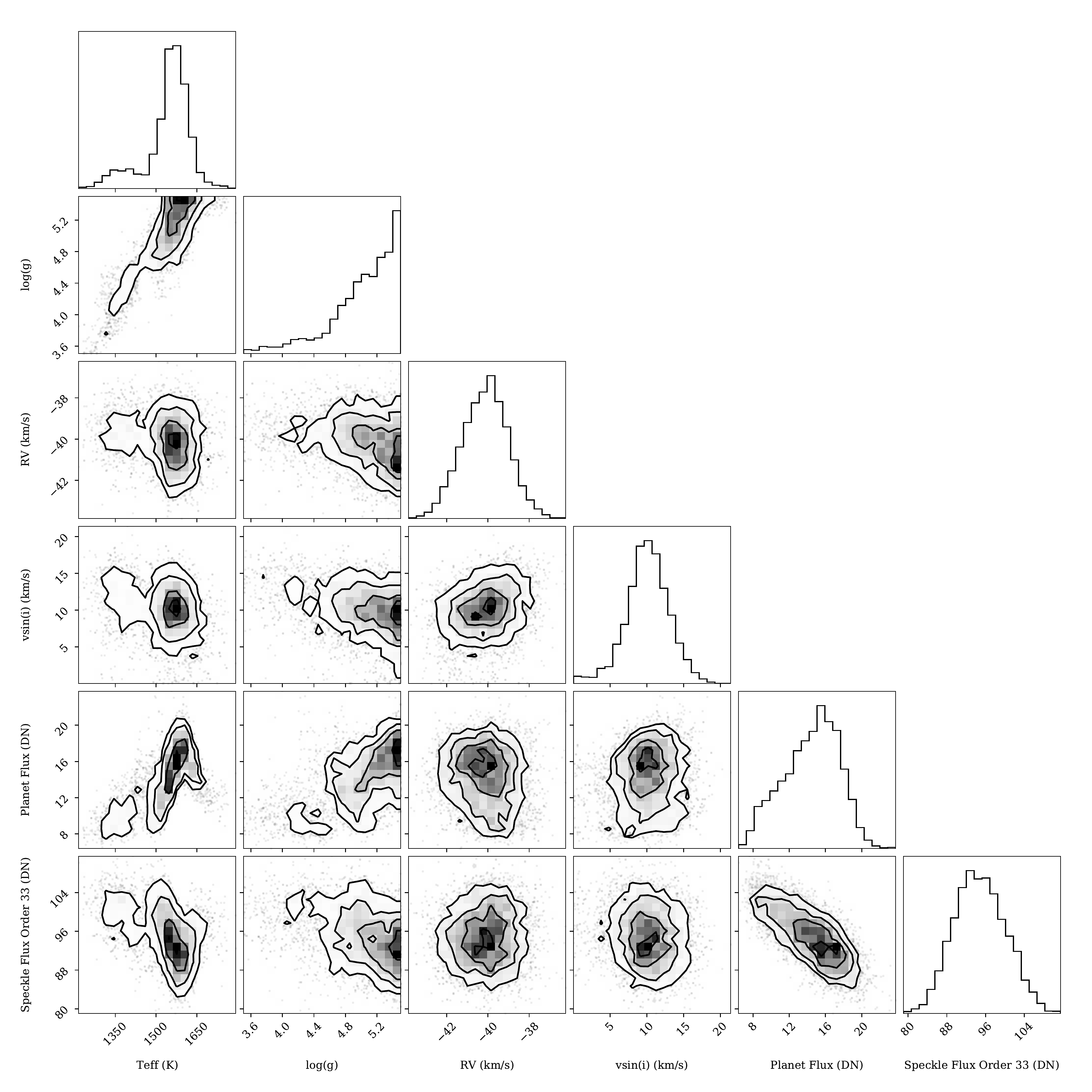}
    \caption{Corner plot for HR 8799 d of the posterior distribution of key parameters in the Full Model fit. \label{fig:corner-d}}
\end{figure}

\begin{figure}
    \centering
    \plotone{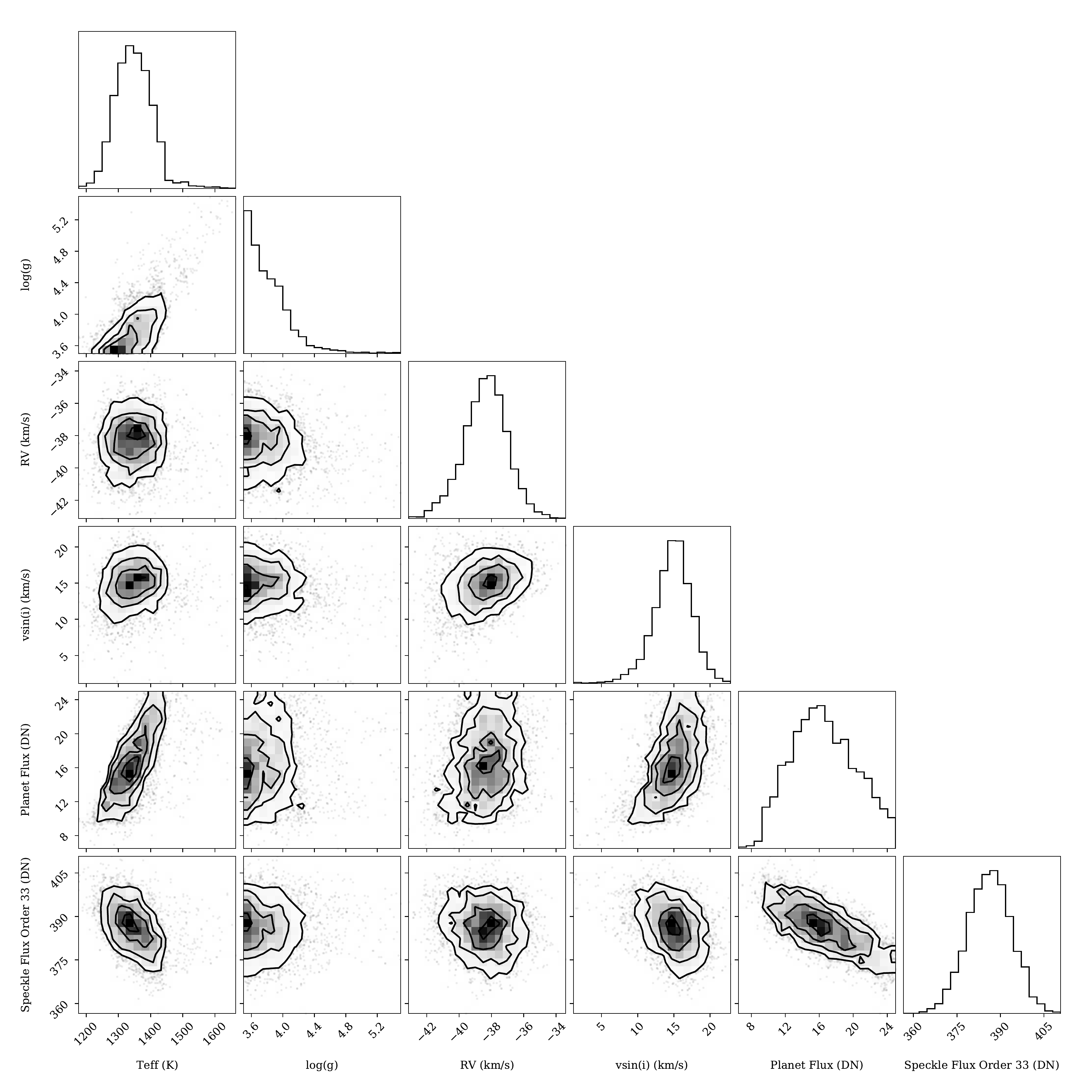}
    \caption{Corner plot for HR 8799 e of the posterior distribution of key parameters in the Full Model fit. \label{fig:corner-e}}
\end{figure}

We present corner plots for the Full Model fits for each planet: HR 8799 b in Figure \ref{fig:corner-b}, HR 8799 c in Figure \ref{fig:corner-c}, HR 8799 d in Figure \ref{fig:corner-d}, and HR 8799 e in Figure \ref{fig:corner-e}. For each planet, we only showed six parameters, which are the main parameters of interest. These included the planet bulk properties $T_\textrm{eff}$, $\log(g)$, planetary RV, $v\sin(i)$, and planetary flux. We also showed the measured speckle flux in Order 33. We did not show the other orders for simplicity as they have the same behavior. We also did not show the error inflation, offset, LSF size scaling, or linear coefficients $c_i$ for combining together the stellar spectra as they are nuisance parameters. The inclusion of all of these parameters would have limited the readability of these plots. 

\section{Forward Model Fit and Residuals}
\label{sec:res-acf}
\begin{figure*}
    \centering
    \includegraphics[width=0.9\textwidth]{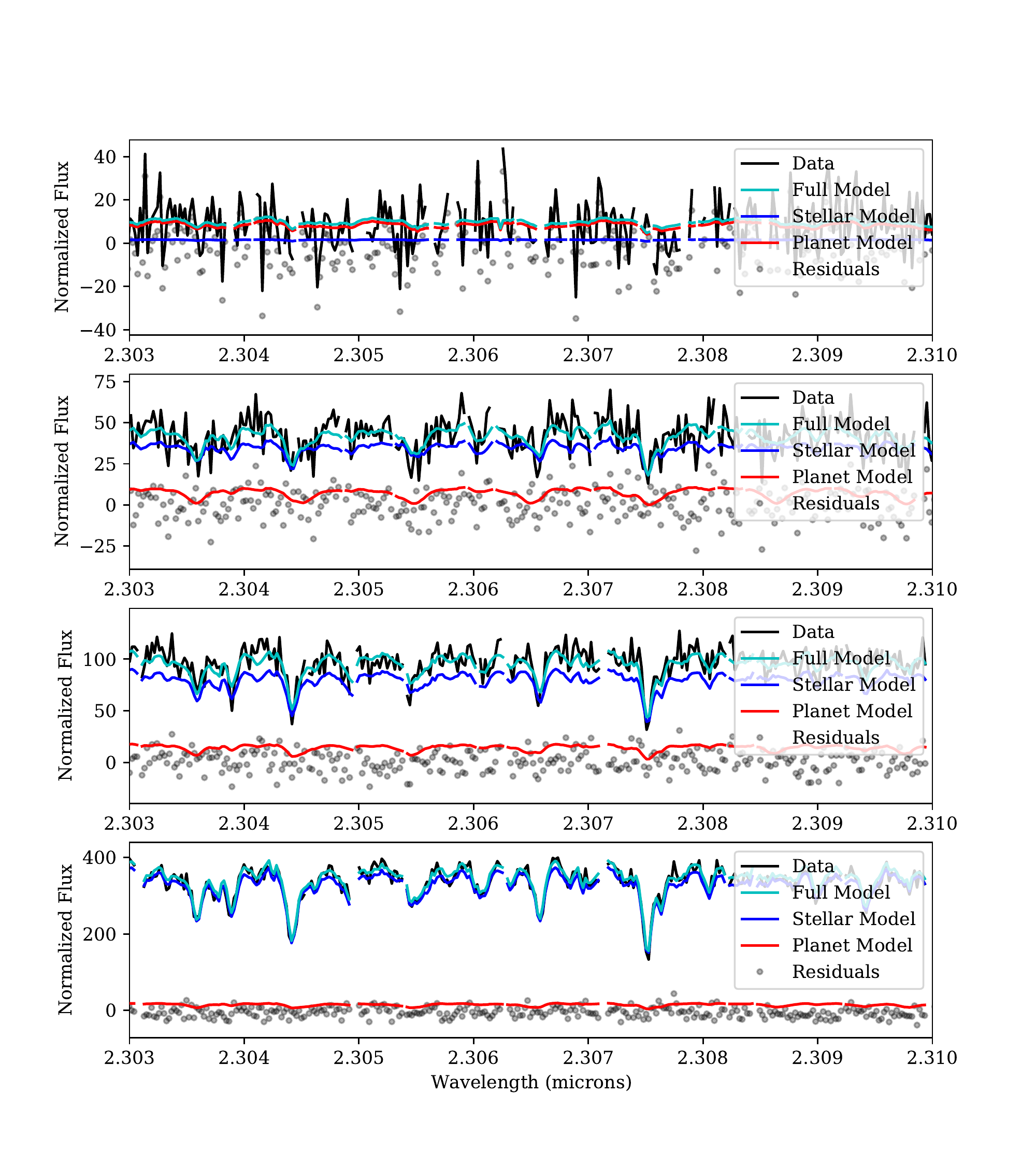}
    \caption{A version of Figure \ref{fig:fm}, except zoomed into a 7 nm chunk of the fit. The gaps in the data and models correspond to channels that were masked due to bad pixels.  }
    \label{fig:zoom-fm}
\end{figure*}

To better to assess the fits in detail, we created a version of Figure \ref{fig:fm} that is zoomed into a 7 nm chunk of the Full Model fit in Order 33 (Figure \ref{fig:zoom-fm}). We can see good agreement between the data and forward model on a channel-to-channel basis. 

Another way to assess the fits are satisfactory is to look for any correlated structure in the residuals. This could be due to model mismatch or correlated noise sources that we had not accounted for. Although we did not visually see any structures in the residuals, the fact there are $\sim$2000 data points per order means we can look for lower levels of correlated residual structure by computing the autocorrelation function (ACF):
\begin{equation}
    \textrm{ACF}(x) = \frac{\sum_i R_i R_{i + x}}{\sum_i R_i R_i}.
\end{equation}
Here, $R$ is the 1D residual vector of length equal to the number of data points in the order, $x$ is the lag in pixels (lag of zero results in $\textrm{ACF} \equiv 1$), and $\sum_i$ iterates over all elements of $R$. For an infinitely long sequence of uncorrelated Gaussian noise, the ACF is the Kronecker delta function. Any correlated structure in the residuals would result in non-zero values in the wings of the ACF. 

\begin{figure}
    \centering
    \includegraphics[width=0.95\textwidth]{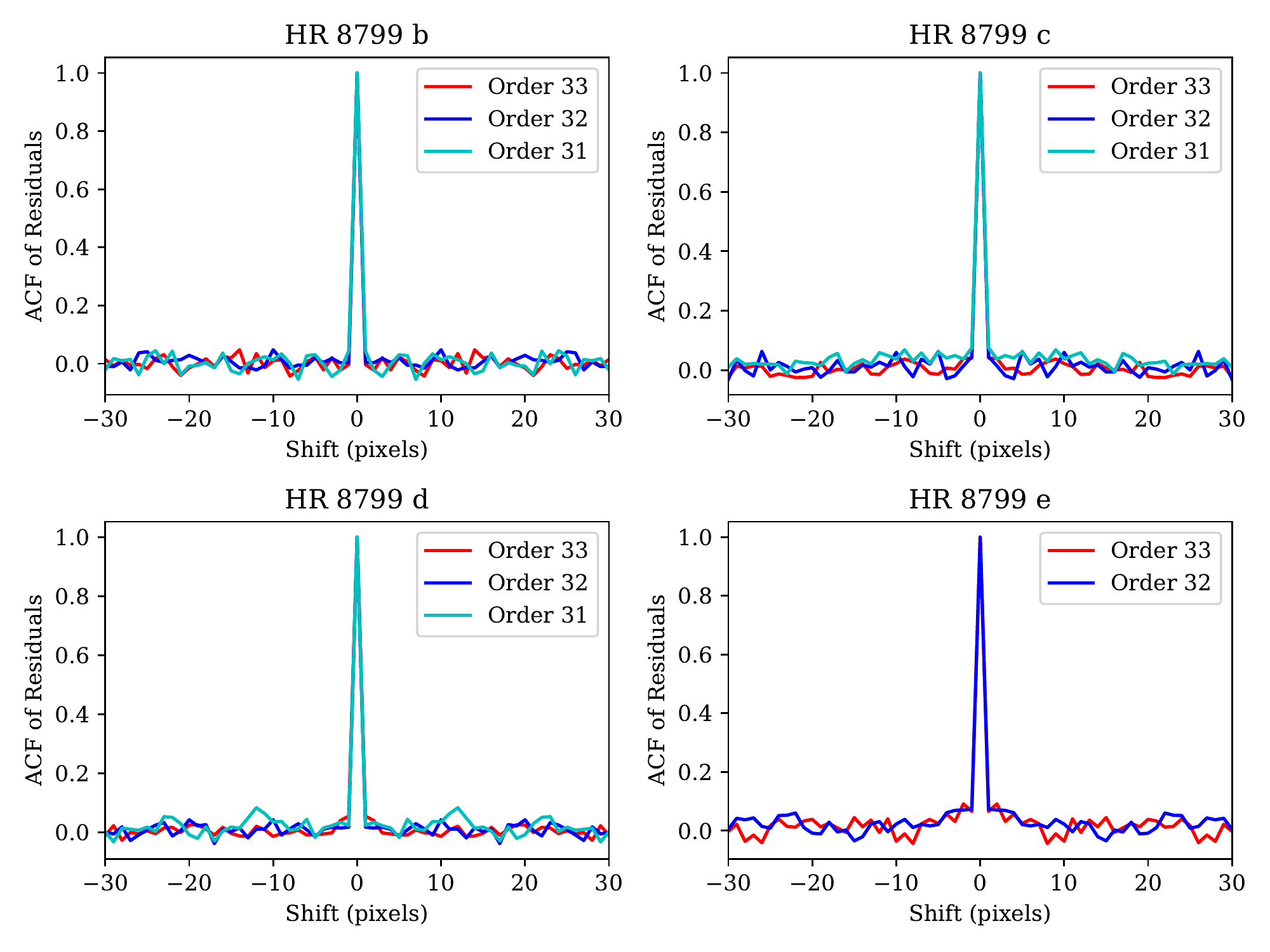}
    \caption{ACF of the fit residuals for the best-fitting model for each planet. For each planet, all of the orders we fit to are plotted. The lack of discernible power in the wings of ACF for each order indicates we are dominated by uncorrelated noise. \label{fig:res-acf}}
\end{figure}

From the fits in Section \ref{sec:fm_fits}, we computed the fit residuals for the parameter set from the Full Model fit of each planet with highest likelihood. For each echelle order used, we computed the ACF of these residuals and plotted them in Figure \ref{fig:res-acf}. From visual inspection of the ACFs, we see that they are almost entirely consistent with a delta function. For HR 8799 e, we saw some power in the wings indicating small correlated structures in the residual ($\leq$ 5\% of the total scatter in the residuals).

\end{appendix}

\end{CJK*}
\end{document}